\documentclass[aps,nofootinbib,preprintnumbers,showpacs,prd,twocolumn,superscriptaddress]{revtex4-2}
\usepackage{array}
\usepackage{booktabs}
\usepackage{subfigure}
\usepackage{capt-of}
\usepackage{graphicx}
\usepackage{lipsum}
\usepackage{epstopdf}
\usepackage{amsmath}
\usepackage{amssymb}
\usepackage{color,xcolor}
\usepackage[bookmarks=false]{hyperref}
\hypersetup{colorlinks=true,citecolor=green,linkcolor=blue,urlcolor=blue,pdfstartview=FitH,bookmarksopen=true}

\begin{document}
	
\title{Wigner distributions of sea quarks in the light-cone quark model}
	
\author{Xiaoyan Luan}
\author{Zhun Lu}\email[]{zhunlu@seu.edu.cn}
\affiliation{School of Physics, Southeast University, Nanjing 211189, China}

\begin{abstract}
		
We investigate the Wigner distributions of $\bar{u}$ and $\bar{d}$ quarks in a proton using the overlap representation within the light cone formalism. Using the light cone wave functions which are obtained from the baryon-meson fluctuation model in terms of the $|q\bar{q}B\rangle$ Fock states, we calculate the Wigner distributions for the unpolarized/longitudinally polarized sea quark in an unpolarized/longitudinally polarized proton. The Wigner distributions can be obtained through a Fourier transform on the generalized transverse-momentum dependent parton distributions (GTMDs). We also calculate the GTMDs of $\bar{u}$ and $\bar{d}$ quarks in the intermediate step. Numerical results for the Wigner distributions of $\bar{u}$ and $\bar{d}$ quarks in transverse momentum space, impact parameter space and the mixed plane are presented. We also study the orbital angular momentum and the spin-orbit correlations of the sea quarks.
\end{abstract}
\maketitle
	
\section{Introduction}\label{Sec1}

Understanding the internal structures of hadrons in terms of quarks and gluon is one of the main goals of QCD and hadronic physics.
In order to describe the inclusive process with one hadron in the initial state,
the parton distribution function (PDF) $f_{i/h}(x)$, which represents the probability density of parton $i$ in a hadron $h$ with the longitudinal momentum fraction $x$, was first introduced by Feynman~\cite{Feynman:1969ej}. 
Although very successful, PDFs can only describe partonic structure of hadrons in one dimension.
In the last three decades, a much more comprehensive picture on the nucleon structure has been developed~\cite{Kotzinian:1994dv,
Mulders:1995dh,Boer:1997nt,Goeke:2005hb,Bacchetta:2006tn}, and the transverse momentum dependent distributions (TMDs) play the central role.
TMDs $f(x,\bm k_\perp)$ not only depend on the longitudinal momentum fraction $x$, but also depend on the parton transverse momentum $\bm k_\perp$ with respect to the hadron, therefore they allow a three-dimensional description of parton structure in momentum space.
TMD distributions naturally enter the description of semi-inclusive deep inelastic scattering and Drell-Yan process in which two hadrons are involved.

Apart from TMDs, in the off-forward region a new type of nucleon structure--the so-call generalized parton distributions (GPDs)~\cite{Muller:1994ses,Ji:1996nm,Radyushkin:1997ki,Mueller:1998fv,
Goeke:2001tz,Diehl:2003ny,Ji:2004gf,
Belitsky:2005qn,Boffi:2007yc}--emerge.
It is the extension of the ordinary PDF from the forward scattering region to the off-forward scattering region.
Therefore, GPDs are natural observables appearing in various exclusive processes in which the target receives a recoil momentum $\Delta$, such as the deeply virtual Compton scattering (DVCS) $\gamma^* h(p) \rightarrow\gamma h (p^\prime)$ and the hard exclusive production of meson $\gamma^* h_1(p) \rightarrow M h_2(p^\prime)$.
Except $x$, GPDs also depend on the momentum transfer squared $t=\Delta^2$ and the longitudinal fraction $\xi=\Delta^+ /P^+$ of the transferred momentum.
Particularly, Fourier transforming GPDs with respect to the transverse component of $\Delta$ yields the impact-parameter dependent distributions (IPDs)~\cite{Burkardt:2000za,Burkardt:2002hr,Diehl:2002he}, $f(x,\bm b_\perp^2)$, with $b_\perp$ the impact parameter conjugate to $\Delta_\perp$. 
IPDs thus provide useful information on the parton tomography in hadrons: distributions of parton in the transverse coordinate space at a fixed $x$.

A more fundamental understanding of the partonic structure of the nucleon can be gained by combining the distributions in momentum space and in position space.
For this purpose, the Wigner distributions of quarks and gluons inside the nucleon~\cite{Ji:2003ak,Belitsky:2003nz} have been proposed and has been extensively studied in recent years. 
The original Wigner distributions for the nucleon are six-dimensional phase-space distributions, which provides a joint transverse momentum space (3D) and transverse position space (3D) about partons in the nucleon. 
Therefore, they encode far more information on the partonic structure of the nucleon than the standard parton distribution functions do. 
A very useful phase-space distribution for describing a fast moving hadron (or in the infinite-momentum frame) is the five-dimensional Wigner distribution~\cite{Lorce:2011kd}, denoted by $W(x,\bm k_\perp, \bm b_\perp)$.
On the one hand, the Wigner distributions reduce to the IPDs  after integrating over transverse momenta.
On the other hand, they reduce to TMDs after integrating over the transverse impact parameters. 
The Wigner distributions cannot be directly measured because of the uncertainty principle which presents the position and momentum of a quantum-mechanical system cannot be simultaneously determined. 
Furthermore the Wigner distributions are usually recognized as quasi-distribution functions which are in general not positive definite and have no probability interpretations. 
However, after integrating over several variables, a reduced Wigner distribution can become positive definite. 

Through Fourier transformations, the Wigner distributions are related to the generalized transverse momentum dependent parton distributions (GTMDs), which are often considered as the ``mother distribution" for TMDs and the GPDs~\cite{Belitsky:2005qn,Ji:2003ak,Belitsky:2003nz}. 
GTMDs are functions of the light-cone momentum fraction, the transverse momentum of the parton as well as the transverse momentum transfer to the nucleon, and are obtained from the generalized parton correlation functions (GPCFs)~\cite{Meissner:2008ay,Meissner:2009ww}  by integrating over the minus component of the parton momentum.
Furthermore, the orbital angular momentum (OAM) of a parton as well as the spin-orbit correlations can be extracted from the Wigner distributions by taking the phase-space average~\cite{Lorce:2011kd,Chakrabarti:2016yuw,Chakrabarti:2017teq}.

In recent years, the Wigner distributions of the valence quarks~\cite{Lorce:2011kd,Lorce:2011ni,Mukherjee:2014nya,More:2017zqq,Liu:2014vwa,
Liu:2015eqa,Miller:2014vla,Muller:2014tqa,Chakrabarti:2016yuw,Chakrabarti:2017teq,
Hagiwara:2014iya,Ma:2018ysi,Kaur:2019kpi} and gluon~\cite{Mukkherjee:2015phf,More:2017zqp,Hagiwara:2016kam,More:2017foj} in the nucleon have been studied in various models. 
In Refs.~\cite{Lorce:2011kd,Lorce:2011ni}, the five-dimensional Wigner distributions $W(x,\bm k_\perp, \bm b_\perp)$ were calculated in the light-cone chiral quark soliton model and the light-cone constituent quark model. 
Besides, the light-front dressed quark model~\cite{Mukherjee:2014nya,Mukkherjee:2015phf,More:2017zqq}, the spectator (diquark) model~\cite{Liu:2014vwa,Liu:2015eqa,Miller:2014vla,Muller:2014tqa,
Miller:2014vla,Muller:2014tqa},the light-cone quark-diquark model~\cite{Chakrabarti:2016yuw,Chakrabarti:2017teq} were applied to calculate the Wigner distribution of the proton.
The light-cone quark model~\cite{Ma:2018ysi} was also used to calculate the Wigner distributions of the pion meson. 
However, the Wigner distribution of the sea quarks are less studied compared to those of the quarks and gluon. 
	
In this paper, we apply the light-cone quark model to calculate the five-dimensional Wigner distributions $\rho\left(x, \boldsymbol{b}_T, \boldsymbol{k}_T\right)$ of the $\bar{u}$ and $\bar{d}$ quarks.  
As pointed out in Ref.~\cite{Lorce:2011kd}, the light-cone formalism is a suitable approach for studying Wigner distributions, since in leading-twist the Wigner distributions can be expressed as the overlap integration of hadronic light-cone wavefunctions~\cite{Brodsky:1997de}.
To generate the sea quark degree freedom, we applied the meson-baryon fluctuation model proposed in Refs.~\cite{Brodsky:1996hc}.
In this model, the proton has the possibility to fluctuate into a composite state with a meson $M$ and a baryon $B$, and consequently the meson $M$ contains the $q\bar{q}$ Fock states.
The light-cone wave functions (LCWFs) of the proton thus may be derived in terms of the $|q\bar{q}B\rangle$ Fock states, as calculated in Ref.~\cite{Luan:2022fjc}. 
The expressions of the Wigner distributions of the unpolarized/longitudinally polarized sea quarks in an unpolarized/longitudinally polarized proton ($\rho_{UU}$, $\rho_{LU}$, $\rho_{LL}$, $\rho_{UL}$) can be obtained in the general case within the overlap representation. 
The numerical results of these Wigner distributions are calculated using the LCWFs from the meson-baryon fluctuation model. 
In the intermediate calculation, the expression of the $\bar{u}$ and $\bar{d}$ quarks GTMD are obtained. 
We also study the OAM and the spin-orbit correlations of $\bar{u}$ and $\bar{d}$ quarks using the relation between the Wigner distributions and GTMDs. 
	
The remained part of the paper is organized as follows. 
In Sec. II, we introduce the definition of the Wigner distributions and their connection with GTMDs. 
In Sec. III, we apply the overlap representation to obtain the expressions of the Wigner distributions and GTMDs of the sea quarks.  
In Sec. VI, we present the numerical results of the unpolarized and longitudinally polarized Wigner distributions in the transverse position space, transverse momentum space and mixed space. 
We summarize the paper in Sec. V.
	
\section{ Wigner distributions and GTMDs}\label{Sec2}
	
The Wigner distribution of partons inside a hadron can be defined as the two-dimensional Fourier transforms of the GTMDs. 
In the light-front framework, the five-dimensional Wigner distribution is written as~\cite{Lorce:2011kd,Meissner:2009ww}
\begin{align}
		\rho_{\Lambda^{\prime} \Lambda}^{[\Gamma]}\left(x, \boldsymbol{b}_T, \boldsymbol{k}_T\right)=\int \frac{\mathrm{d}^{2} \boldsymbol{\Delta}_T}{(2 \pi)^{2}} e^{-i \boldsymbol{\Delta}_T \cdot \boldsymbol{b}_T} W_{\Lambda^{\prime} \Lambda}^{[\Gamma]}\left(x, \boldsymbol{\Delta}_T, \boldsymbol{k}_T\right),\label{eq:rholambda}
\end{align}
where $\boldsymbol{\Delta}_T$ is the momentum transfer from the initial state to the final state in the transverse direction, and $\boldsymbol{b}_T$ is the impact parameter in the position space conjugate to $\boldsymbol{\Delta}_T$. 
Similar to the standard quark-quark correlation operator, the generalized correlator  $W_{\Lambda^{\prime} \Lambda}^{[\Gamma]}\left(x, \boldsymbol{\Delta}_T, \boldsymbol{k}_T\right)$ at $\xi=0$ and a fixed light-cone time is defined as:
\begin{align}\label{correlator}
\notag&W_{\Lambda^{\prime} \Lambda}^{[\Gamma]}\left(x, \boldsymbol{\Delta}_T, \boldsymbol{k}_T\right)\\&=\left.\int \frac{d z^{-} d^{2} z_T}{2(2 \pi)^{3}} e^{i k \cdot z}\left\langle p^{\prime}\Lambda^{\prime}\left|\bar{\psi}\left(-\frac{z}{2} \right) \Gamma \mathcal{W} \psi\left(\frac{z}{2} \right)\right| p\Lambda\right\rangle\right|_{z^{+}=0},
\end{align}
where $\mathcal{W}$ is the gauge link connecting the quark fields at positions $-\frac{z}{2}$ and $\frac{z}{2}$ to ensure the color gauge invariance, $p\Lambda $ ($p^{\prime} \Lambda^{\prime}$) are the momenta and helicities of the initial (final) state nucleon, $P=(p+p^{\prime})/2$ is the average four-momentum of the nucleon, $x=k^+/P^+$ is the average fraction of the light-cone momentum carried by the active quark, $\Gamma$ represents the Dirac matrix structure. 
In this work we take $\Gamma=\gamma^{+}, \gamma^{+} \gamma^{5}$.
	
The generalized correlator in Eq.(\ref{correlator}) can be parameterized in terms of GTMDs~\cite{Meissner:2009ww}. 
At leading twist $\Gamma=\gamma^{+}$ or $\gamma^{+} \gamma^{5}$, we have eight GTMDs defined as follows:
\begin{widetext}
\begin{align}\label{W+}
\notag	W_{\Lambda^{\prime} \Lambda}^{[\gamma^{+}]} & =\frac{1}{2 M} \bar{U}\left(p^{\prime}, \Lambda^{\prime}\right)\left[F_{11}+\frac{i \sigma^{i+} k^{i}}{P^{+}} F_{12}+\frac{i \sigma^{i+} \Delta^{i}}{P^{+}} F_{13}+\frac{i \sigma^{i j} k^{i} \Delta^{j}}{M^{2}} F_{14}\right] U(p, \Lambda) \\
& =\left[F_{11}+\frac{i \Lambda \epsilon^{i j} k^{i} \Delta^{j}}{M^{2}} F_{14}\right] \delta_{\Lambda^{\prime} \Lambda}+\left[\frac{\Lambda \Delta^{1}+i \Delta^{2}}{2 M}\left(2 F_{13}-F_{11}\right)+\frac{\Lambda k^{1}+i k^{2}}{M} F_{12}\right] \delta_{-\Lambda^{\prime} \Lambda},\\
\label{W+5}\notag	W_{\Lambda^{\prime} \Lambda}^{[\gamma^{+} \gamma^{5}]} & =\frac{1}{2 M} \bar{U}\left(p^{\prime}, \Lambda^{\prime}\right)\left[-\frac{i \epsilon^{i j} k^{i} \Delta^{j}}{M^{2}} G_{11}+\frac{i \sigma^{i+} \gamma^{5} k^{i}}{P^{+}} G_{12}+\frac{i \sigma^{i+} \gamma^{5} \Delta^{i}}{P^{+}} G_{13}+i \sigma^{+-} \gamma^{5} G_{14}\right] U(p, \Lambda) \\
& =\left[-\frac{i\left(k^{1} \Delta^{2}-k^{2} \Delta^{1}\right)}{M^{2}} G_{11}+\Lambda G_{14}\right] \delta_{\Lambda^{\prime} \Lambda}+\left[\frac{\Delta^{1}+i \Lambda \Delta^{2}}{M}\left(G_{13}+\frac{i \Lambda\left(k^{1} \Delta^{2}-k^{2} \Delta^{1}\right)}{2 M^{2}} G_{11}\right)+\frac{k^{1}+i \Lambda k^{2}}{M} G_{12}\right] \delta_{-\Lambda^{\prime} \Lambda}.
\end{align} 
Here $\varepsilon^{i j}=\varepsilon^{-+i j}$ and $\varepsilon^{0123}=1$. 
The GTMDs $F_{11}$, $F_{12}$, $F_{13}$, $F_{14}$ in Eq.~(\ref{W+}) describe the distribution for the unpolarized quark, while the GTMDs $G_{11}$, $G_{12}$, $G_{13}$, $G_{14}$ in Eq.~(\ref{W+5}) describe those for the longitudinally polarized quark. 
They are considered to be complex functions of the kinematical variables $x, \xi, \boldsymbol{\Delta}_{T}^2, \boldsymbol{k}_{T}\cdot\boldsymbol{\Delta}_{T}, \boldsymbol{k}_T^2$. 
	
Using $+$ ($-$) denote the positive (negative) helicity of the proton, we can obtain the following expressions for the terms in which the proton helicity is not flipped~\cite{Rajan:2017cpx}
\begin{align}
		F_{11}\left(x, \boldsymbol{\Delta}_{T}, \boldsymbol{k}_T\right)&=\frac{1}{2}\left[W_{++}^{[\gamma^{+}]}\left(x, \boldsymbol{\Delta}_{T}, \boldsymbol{k}_T\right)+W_{--}^{[\gamma^{+}]}\left(x, \boldsymbol{\Delta}_{T}, \boldsymbol{k}_T\right)\right], 
		\\
		i\frac{\boldsymbol{k}_T\times\boldsymbol{\Delta}_T}{M^2}F_{14}\left(x, \boldsymbol{\Delta}_{T}, \boldsymbol{k}_T\right)&=\frac{1}{2}\left[W_{++}^{[\gamma^{+}]}\left(x, \boldsymbol{\Delta}_{T}, \boldsymbol{k}_T\right)-W_{--}^{[\gamma^{+}]}\left(x, \boldsymbol{\Delta}_{T}, \boldsymbol{k}_T\right)\right]  ,
		\\
    	-i\frac{\boldsymbol{k}_T\times\boldsymbol{\Delta}_T}{M^2}G_{11}\left(x, \boldsymbol{\Delta}_{T}, \boldsymbol{k}_T\right)&=\frac{1}{2}\left[W_{++}^{[\gamma^{+}\gamma^{5}]}\left(x, \boldsymbol{\Delta}_{T}, \boldsymbol{k}_T\right)+W_{--}^{[\gamma^{+}\gamma^{5}]}\left(x, \boldsymbol{\Delta}_{T}, \boldsymbol{k}_T\right)\right], \\
		G_{14}\left(x, \boldsymbol{\Delta}_{T}, \boldsymbol{k}_T\right)&=\frac{1}{2}\left[W_{++}^{[\gamma^{+}\gamma^{5}]}\left(x, \boldsymbol{\Delta}_{T}, \boldsymbol{k}_T\right)-W_{--}^{[\gamma^{+}\gamma^{5}]}\left(x, \boldsymbol{\Delta}_{T}, \boldsymbol{k}_T\right)\right].    	
\end{align} 
and expressions for the helicity-flipped terms~\cite{Rajan:2017cpx}
\begin{align}
-i\frac{\boldsymbol{k}_{T} \times \boldsymbol{\Delta}_{T}}{M} F_{12}& = \frac{1}{2}\left[(\left(\Delta^{1}-i \Delta^{2}\right) W_{-+}^{[\gamma^{+}]}+\left(\Delta^{1}+i \Delta^{2}\right) W_{+-}^{[\gamma^{+}]}\right], 
		\\
		\frac{\boldsymbol{k}_{T} \cdot \boldsymbol{\Delta}_{T}}{M} F_{12}+\frac{\boldsymbol{\Delta}_{T}^{2}}{2 M}\left(2 F_{13}-F_{11}\right)& = \frac{1}{2}\left[(\left(\Delta^{1}-i \Delta^{2}\right) W_{-+}^{[\gamma^{+}]}-\left(\Delta^{1}+i \Delta^{2}\right) W_{+-}^{[\gamma^{+}]}\right],
		\\
		\frac{\boldsymbol{\Delta}_{T}^{2}}{M} G_{13}+\frac{\boldsymbol{k}_{T} \cdot \boldsymbol{\Delta}_{T}}{M} G_{12}& = \frac{1}{2}\left[(\left(\Delta^{1}-i \Delta^{2}\right) W_{-+}^{[\gamma^{+}\gamma^{5}]}+\left(\Delta^{1}+i \Delta^{2}\right) W_{+-}^{[\gamma^{+}\gamma^{5}]}\right], 
		\\
		\frac{i\left(\boldsymbol{k}_{T} \times \boldsymbol{\Delta}_{T}\right)}{M}\left(\frac{\boldsymbol{\Delta}_{T}^{2}}{2 M^{2}} G_{11}-G_{12}\right) & = \frac{1}{2}\left[\left(\Delta^{1}-i \Delta^{2}\right) W_{-+}^{[\gamma^{+}\gamma^{5}]}-\left(\Delta^{1}+i \Delta^{2}\right) W_{+-}^{[\gamma^{+}\gamma^{5}]}\right] .
	\end{align}
\end{widetext}

Using the notation in Eq.~(\ref{eq:rholambda}), one can define four Wigner distributions. 
The first one is the Wigner distribution of the unpolarized quark in the unpolarized target
\begin{align}
\rho_{U U}\left(x, \boldsymbol{b}_T, \boldsymbol{k}_T\right)=\frac{1}{2}\left[\rho_{++}^{[\gamma^{+}]}\left(x, \boldsymbol{b}_T, \boldsymbol{k}_T\right)+\rho_{--}^{[\gamma^{+}]}\left(x, \boldsymbol{b}_T, \boldsymbol{k}_T\right)\right],
\end{align}
the second one is the Wigner distribution of the unpolarized quark in the longitudinally polarized target
\begin{align}
\rho_{L U}\left(x, \boldsymbol{b}_T, \boldsymbol{k}_T\right)=\frac{1}{2}\left[\rho_{++}^{[\gamma^{+}]}\left(x, \boldsymbol{b}_T, \boldsymbol{k}_T\right)-\rho_{--}^{[\gamma^{+}]}\left(x, \boldsymbol{b}_T, \boldsymbol{k}_T\right)\right],
\end{align}
the third one is the Wigner distribution of the longitudinally polarized quark in the unpolarized target
\begin{align}
\rho_{U L}\left(x, \boldsymbol{b}_T, \boldsymbol{k}_T\right)&=\frac{1}{2}\left[\rho_{++}^{[\gamma^{+}\gamma^{5}]}\left(x, \boldsymbol{b}_T, \boldsymbol{k}_T\right)\right.\nonumber\\
&\left.+\rho_{--}^{[\gamma^{+}\gamma^{5}]}\left(x, \boldsymbol{b}_T, \boldsymbol{k}_T\right)\right],
\end{align}
and	the last one is the Wigner distribution of the longitudinally polarized quark in the longitudinally polarized target
\begin{align}
\rho_{L L}\left(x, \boldsymbol{b}_T, \boldsymbol{k}_T\right)&=\frac{1}{2}\left[\rho_{++}^{[\gamma^{+}\gamma^{5}]}\left(x, \boldsymbol{b}_T, \boldsymbol{k}_T\right)\right.\nonumber\\
&\left.-\rho_{--}^{[\gamma^{+}\gamma^{5}]}\left(x, \boldsymbol{b}_T, \boldsymbol{k}_T\right)\right].
\end{align}
	
Finally, these Wigner distributions can be obtained from the GTMDs
\begin{align}
\rho_{U U}\left(x, \boldsymbol{b}_T, \boldsymbol{k}_T\right)&=\mathcal{F}_{11}(x,0, \boldsymbol{k}_T^2,\boldsymbol{k}_T\cdot\boldsymbol{b}_T, \boldsymbol{b}_T^2), \\
\rho_{L U}\left(x, \boldsymbol{b}_T, \boldsymbol{k}_T\right)&=-\frac{1}{M^2}\epsilon^{ij}_Tk_T^i\frac{\partial}{\partial b_T^j}\mathcal{F}_{14}(x,0, \boldsymbol{k}_T^2,\boldsymbol{k}_T\cdot\boldsymbol{b}_T, \boldsymbol{b}_T^2),\label{F14}  \\
\rho_{U L}\left(x, \boldsymbol{b}_T, \boldsymbol{k}_T\right)&=\frac{1}{M^2}\epsilon^{ij}_Tk_T^i\frac{\partial}{\partial b_T^j}\mathcal{G}_{11}(x,0, \boldsymbol{k}_T^2,\boldsymbol{k}_T\cdot\boldsymbol{b}_T, \boldsymbol{b}_T^2), \label{G14} \\
\rho_{L L}\left(x, \boldsymbol{b}_T, \boldsymbol{k}_T\right)&=\mathcal{G}_{14}(x,0, \boldsymbol{k}_T^2,\boldsymbol{k}_T\cdot\boldsymbol{b}_T, \boldsymbol{b}_T^2),
\end{align}
where $\mathcal{X}$ is the Fourier transform of the corresponding GTMD $X$
\begin{align}
\mathcal{X}\left(x, \boldsymbol{b}_T, \boldsymbol{k}_T\right)=\int \frac{\mathrm{d}^{2} \boldsymbol{\Delta}_T}{(2 \pi)^{2}} e^{-i \boldsymbol{\Delta}_T \cdot \boldsymbol{b}_T} X\left(x, \boldsymbol{\Delta}_T, \boldsymbol{k}_T\right).
\end{align}
	
\section{sea quark wigner distributions of the proton in the overlap representation}

In this section we present the calculation on the sea quark Wigner distributions of the proton in the light-cone model using the overlap representation. 
The light-cone formalism has been widely recognized as a convenient way to calculate the parton distribution functions of nucleon and meson~\cite{Lepage:1980fj}. 
Within the light-cone approach, a hadronic composite state can be expressed as LCWFs on the Fock-state basis. 
The overlap representation has also been used to study various form factors of the hadrons~\cite{Brodsky:2000ii,Xiao:2003wf}, anomalous magnetic moment of the nucleon~\cite{Brodsky:2000ii}, TMDs~\cite{Bacchetta:2008af,Lu:2006kt} as well as GPDs~\cite{Muller:2014tqa,Brodsky:2000xy,Burkardt:2003je,Luan:2023lmt}. 
Recently, the overlap representation of LCWFs has also been applied to calculate the quark Wigner distributions~\cite{Ma:2018ysi,Kaur:2020par}. 
Here we extend light-cone formalism to calculate the Wigner distributions and GTMDs of the sea quark. 
	
The baryon-meson fluctuation model~\cite{Brodsky:1996hc} is adopted to generate the degree of freedom of sea quark, in which the proton can fluctuate to a composite system formed by a meson $M$ and a baryon $B$, where the meson is composed in terms of $q\bar{q}$.
\begin{align}\label{fock_state}
|p\rangle\to| M B\rangle\to|q\bar{q}B\rangle.
\end{align}
In our work, we consider the fluctuation $|p\rangle \to |\pi^+ n\rangle $ and $|p\rangle \to |\pi^- \Delta^{++}\rangle $. 
The corresponding LCWFs have the form which have been derived in Ref.~\cite{Luan:2022fjc}
\begin{align}\label{LCWFs} \psi^{\lambda_N}_{{\lambda_B}{\lambda_q}{\lambda_{\bar{q}}}}(x,y,\boldsymbol{k}_T,\boldsymbol{r}_T)		=&\psi^{\lambda_N}_{\lambda_B}(y,\boldsymbol{r}_T)\psi_{{\lambda_q}{\lambda_{\bar{q}}}}
(x,y,\boldsymbol{k}_T,\boldsymbol{r}_T),
\end{align}
where $\psi^{\lambda_N}_{\lambda_B}(y,r_T)$ can be viewed as the wave function of the nucleon in terms $\pi B$ components, and $\psi_{{\lambda_q}{\lambda_{\bar{q}}}}
(x,y,\boldsymbol{k}_T,\boldsymbol{r}_T)$ is the pion wave function in terms of $q \bar{q}$ components. The indices $\lambda_N$, $\lambda_B$, $\lambda_q$, $\lambda_{\bar{q}}$ denote the helicity of the proton, the baryon, the quark and the sea quark, respectively. 
$x$ and $y$ represent their light-cone momentum fractions, $\boldsymbol{k}_T$ and $\boldsymbol{r}_T$ denote the transverse momenta of the antiquark and the meson. 
For the former one of Eq.~(\ref{LCWFs}), they have the expression:
\begin{align}\label{former}
		\notag\psi^+_+(y,\boldsymbol{r}_T)&=\frac{M_B-(1-y)M}{\sqrt{1-y}}\phi_1, \\
		\notag\psi^+_-(y,\boldsymbol{r}_T)&=\frac{r_1+ir_2}{\sqrt{1-y}}\phi_1, \\
		\notag\psi^-_+(y,\boldsymbol{r}_T)&=\frac{r_1-ir_2}{\sqrt{1-y}}\phi_1 , \\
		\psi^-_-(y,\boldsymbol{r}_T)&=\frac{(1-y)M-M_B}{\sqrt{1-y}}\phi_1,
\end{align}	
Here, $M$ and $M_B$ are the masses of proton and baryon, respectively. $\phi_1$ is the momentum space wave function of the bayron-meson Fock state 
\begin{align}		\phi_1(y,\boldsymbol{r}_T)=-\frac{g(r^2)\sqrt{y(1-y)}}{\boldsymbol{r}_T^2+L_1^2(m_\pi^2)},
\end{align}
where $m_\pi$ is the mass of $\pi$ meson, $g(r^2)$ is the form factor for the coupling of the nucleon-pion meson-baryon vertex, and
\begin{align}
		L_1^2({m_\pi^2})=yM_B^2+(1-y){m_\pi^2}-y(1-y)M^2.
\end{align}
The latter one of Eq.~(\ref{LCWFs}) have the following expressions:
\begin{align}\label{later}		\notag\psi{_+}{_+}(x,y,\boldsymbol{k}_T,\boldsymbol{r}_T)&=\frac{my}{\sqrt{x(y-x)}}\phi_2,\\ \notag\psi{_+}{_-}(x,y,\boldsymbol{k}_T,\boldsymbol{r}_T)&=\frac{y(k_1-ik_2)-x(r_1-ir_2)}{\sqrt{x(y-x)}}\phi_2, \\ \notag\psi{_-}{_+}(x,y,\boldsymbol{k}_T,\boldsymbol{r}_T)&=\frac{y(k_1+ik_2)-x(r_1+ir_2)}{\sqrt{x(y-x)}}\phi_2, \\		\psi{_-}{_-}(x,y,\boldsymbol{k}_T,\boldsymbol{r}_T)&=\frac{-my}{\sqrt{x(y-x)}}\phi_2,
\end{align} 
Here, $m$ is the mass of quark and the sea quark. 
Again, $\phi_2(x,y,\boldsymbol{k}_T,\boldsymbol{r}_T)$ is the momentum space wave function of the $|q\bar{q}\rangle$ Fock state 
\begin{align}	     	\phi_2(x,y,\boldsymbol{k}_T,\boldsymbol{r}_T)=
-\frac{g(k^2)\sqrt{\frac{x}{y}(1-\frac{x}{y})}}{(\boldsymbol{k}_T
-\frac{x}{y}\boldsymbol{r}_T)^2+L_2^2(m^2)},
\end{align}
$g(k^2)$ is the form factor for the coupling of the pion meson-quark-sea quark vertex, and
\begin{align}		
L_2^2(m^2)=\frac{x}{y}m^2+\left(1-\frac{x}{y}\right)m^2
-\frac{x}{y}\left(1-\frac{x}{y}\right){m_\pi}^2.
\end{align}
For the form factors $g(r^2)$ and $g(k^2)$, we adopt the dipolar form
\begin{align}
g(r^2)&=-g_1(1-y)\frac{\boldsymbol{r}_T^2+L_1^2(m_\pi^2)}
{[\boldsymbol{r}_T^2+L_1^2(\Lambda^2_\pi)]^2},\label{eq15}	\\		g(k^2)&=-g_2(1-\frac{x}{y})\frac{(\boldsymbol{k}_T-\frac{x}{y}\boldsymbol{r}_T)^2+L_2^2(m^2)}{[(\boldsymbol{k}_T-\frac{x}{y}\boldsymbol{r}_T)^2+L_2^2(\Lambda^2_{\bar{q}})]^2}.\label{eq16}	
\end{align}
	
In the overlap representation, the leading-twist generalized correlator can be expressed as
\begin{align}\label{W1}
\notag&W_{\Lambda^{\prime} \Lambda}^{[\gamma^{+}]}(x,\boldsymbol{k}_T,\boldsymbol{\Delta}_T)
=\int\frac{d^2\boldsymbol{r}_T}{16\pi^{3}}\int_{x}^{1}\frac{dy}{y}\\
&\times\sum_{\{\lambda\}}\left(\psi_{\lambda_{B}\lambda_{q}
\lambda_{\bar{q}}}^{\Lambda^{\prime}\star}\left(x^{out}, y^{out}, \boldsymbol{k}_T^{out},\boldsymbol{r}_T^{out}\right)\right.  \nonumber\\ &\left.\times\psi_{\lambda_{B}\lambda_{q}\lambda_{\bar{q}}}^{\Lambda}\left(x^{in}, y^{in}, \boldsymbol{k}_T^{in},\boldsymbol{r}_T^{in}\right)\right),\\
\label{W2}\notag	
&W_{\Lambda^{\prime} \Lambda}^{[\gamma^{+}\gamma^{5}]}(x,\boldsymbol{k}_T,\boldsymbol{\Delta}_T)
=\int\frac{d^2\boldsymbol{r}_T}{16\pi^{3}}\int_{x}^{1}\frac{dy}{y} \textrm{sign}(\lambda_{\bar{q}})\\
&\times\sum_{\{\lambda\}}\left(\psi_{\lambda_{B}\lambda_{q}
\lambda_{\bar{q}}}^{\Lambda^{\prime}\star}\left(x^{out}, y^{out}, \boldsymbol{k}_T^{out},\boldsymbol{r}_T^{out}\right)  \right.\nonumber\\ 
&\times \psi_{\lambda_{B}\lambda_{q}\lambda_{\bar{q}}}^{\Lambda}\left.\left(x^{in}, y^{in}, \boldsymbol{k}_T^{in},\boldsymbol{r}_T^{in}\right)\right).
\end{align}
where $\{\lambda\}={\lambda_B, \lambda_q, \lambda_{\bar{q}}}$. 
The momentum fractions for the final-state and initial-state antiquark $\bar{q}$ and the meson can be expressed as
\begin{align}
\nonumber x^{out}=\frac{x-\xi}{1-\xi} \quad x^{in}=\frac{x+\xi}{1+\xi}  \\
y^{out}=\frac{y-\xi}{1-\xi} \quad y^{in}=\frac{y+\xi}{1+\xi},
\end{align}	
and
\begin{align}		\nonumber\boldsymbol{k}^{out}_T&=\boldsymbol{k}_T-\frac{1}{2}(1-x^{out})\boldsymbol{\Delta}_T
\\
\boldsymbol{k}^{in}_T&=\boldsymbol{k}_T+\frac{1}{2}(1-x^{in})\boldsymbol{\Delta}_T,
\end{align}	
are the expressions for transverse momenta for the antiquark.
Finally, the transverse momenta for the spectator baryon and the quark has the form: 
\begin{align}
\nonumber  -\boldsymbol{r}^{out}_T&=-\boldsymbol{r}_T+\frac{1}{2}(1-y)\boldsymbol{\Delta}_T\\
\nonumber  -\boldsymbol{r}^{in}_T&=-\boldsymbol{r}_T-\frac{1}{2}(1-y)\boldsymbol{\Delta}_T\\
\nonumber(\boldsymbol{r}_T-\boldsymbol{k}_T)^{out}&=	(\boldsymbol{r}_T-\boldsymbol{k}_T)+\frac{1}{2}(y-x)\boldsymbol{\Delta}_T \\
(\boldsymbol{r}_T-\boldsymbol{k}_T)^{in}&=	(\boldsymbol{r}_T-\boldsymbol{k}_T)-\frac{1}{2}(y-x)\boldsymbol{\Delta}_T,
\end{align}	
	
Using the LCWFs in Eqs.~(\ref{former}-\ref{later}) and the overlap representation for the generalized correlator in Eqs.~(\ref{W1}-\ref{W2}), we get the results of the GTMDs  of the sea quark in proton at $\xi=0$
\begin{widetext}
\begin{align}
\notag
F_{11}\left(x, \boldsymbol{\Delta}_{T}, \boldsymbol{k}_T\right)&=\frac{g_1^2g_2^2}{8\pi^3}\int_{x}^{1}\frac{dy}{y}\int d^2\boldsymbol{r}_T		\\&\frac{y(1-y)^2(1-\frac{x}{y})^2\left[\boldsymbol{r}_T^2-\frac{1}{4}(1-y)^2
\boldsymbol{\Delta}_T^2+[M_B-(1-y)M]^2\right]\left[(\boldsymbol{k}_T
-\frac{x}{y}\boldsymbol{r}_T)^2-\frac{1}{4}(1-\frac{x}{y})^2
\boldsymbol{\Delta}_T^2+m^2\right]}{D_1(y,\boldsymbol{r}_T,\boldsymbol{\Delta}_T) D_2(\frac{x}{y},\boldsymbol{k}_T-\frac{x}{y}\boldsymbol{r}_T,\boldsymbol{\Delta}_T)},
\\
F_{12}\left(x, \boldsymbol{\Delta}_{T}, \boldsymbol{k}_T\right)&=0,\\
\notag 
F_{13}\left(x, \boldsymbol{\Delta}_{T}, \boldsymbol{k}_T\right)&=\frac{g_1^2g_2^2}{16\pi^3}\int_{x}^{1}\frac{dy}{y}\int d^2\boldsymbol{r}_Ty(1-y)^2(1-\frac{x}{y})^2		\frac{\left[(\boldsymbol{k}_T-\frac{x}{y}\boldsymbol{r}_T)^2
-\frac{1}{4}(1-\frac{x}{y})^2\boldsymbol{\Delta}_T^2+m^2\right]}{ D_2(\frac{x}{y},\boldsymbol{k}_T-\frac{x}{y}\boldsymbol{r}_T,\boldsymbol{\Delta}_T)}	\\&\times\frac{\left[2M(1-y)[(1-y)M-M_B]+\boldsymbol{r}_T^2
-\frac{1}{4}(1-y)^2\boldsymbol{\Delta}_T^2+[M_B-(1-y)M]^2\right]}
{D_1(y,\boldsymbol{r}_T,\boldsymbol{\Delta}_T) },\\
\notag 
F_{14}\left(x, \boldsymbol{\Delta}_{T}, \boldsymbol{k}_T\right)&=\frac{g_1^2g_2^2}{8\pi^3}\int_{x}^{1}\frac{dy}{y}\int d^2\boldsymbol{r}_T\frac{y(1-y)^3(1-\frac{x}{y})^2M^2\left[(\boldsymbol{k}_T
-\frac{x}{y}\boldsymbol{r}_T)^2-\frac{1}{4}(1-\frac{x}{y})^2
\boldsymbol{\Delta}_T^2+m^2\right]}{D_1(y,\boldsymbol{r}_T,\boldsymbol{\Delta}_T) D_2(\frac{x}{y},\boldsymbol{k}_T-\frac{x}{y}\boldsymbol{r}_T,\boldsymbol{\Delta}_T)}
\cdot\frac{\boldsymbol{r}_T\times\boldsymbol{\Delta}_T}{\boldsymbol{k}_T\times
\boldsymbol{\Delta}_T},\\
\notag
G_{11}\left(x, \boldsymbol{\Delta}_{T}, \boldsymbol{k}_T\right)&=-\frac{g_1^2g_2^2}{8\pi^3}\int_{x}^{1}\frac{dy}{y}\int d^2\boldsymbol{r}_T\\
&\frac{y(1-y)^2(1-\frac{x}{y})^3M^2\left[\boldsymbol{r}_T^2-\frac{1}{4}(1-y)^2
\boldsymbol{\Delta}_T^2+[M_B-(1-y)M]^2\right]}{D_1(y,\boldsymbol{r}_T,\boldsymbol{\Delta}_T) D_2(\frac{x}{y},\boldsymbol{k}_T-\frac{x}{y}\boldsymbol{r}_T,\boldsymbol{\Delta}_T)}
\cdot\frac{(\boldsymbol{k}_T-\frac{x}{y}\boldsymbol{r}_T)\times
\boldsymbol{\Delta}_T}{\boldsymbol{k}_T\times\boldsymbol{\Delta}_T},\\
\notag	
	G_{12}\left(x, \boldsymbol{\Delta}_{T}, \boldsymbol{k}_T\right)&=\frac{g_1^2g_2^2}{16\pi^3}\int_{x}^{1}\frac{dy}{y}\int d^2\boldsymbol{r}_Ty(1-y)^2(1-\frac{x}{y})^3\\
&\frac{\boldsymbol{\Delta}_T^2\left[2M(1-y)(M_B-(1-y)M)-\boldsymbol{r}_T^2+\frac{1}{4}(1-y)^2\boldsymbol{\Delta}_T^2-[M_B-(1-y)M]^2\right]}{D_1(y,\boldsymbol{r}_T,\boldsymbol{\Delta}_T) D_2(\frac{x}{y},\boldsymbol{k}_T-\frac{x}{y}\boldsymbol{r}_T,\boldsymbol{\Delta}_T)}\cdot\frac{(\boldsymbol{k}_T-\frac{x}{y}\boldsymbol{r}_T)\times\boldsymbol{\Delta}_T}{\boldsymbol{k}_T\times\boldsymbol{\Delta}_T},
\\
\notag
G_{13}\left(x, \boldsymbol{\Delta}_{T}, \boldsymbol{k}_T\right)&=-\frac{g_1^2g_2^2}{16\pi^3}\int_{x}^{1}\frac{dy}{y}\int d^2\boldsymbol{r}_Ty(1-y)^2(1-\frac{x}{y})^3\\
&\frac{\boldsymbol{k}_T \cdot\boldsymbol{\Delta}_T\left[2M(1-y)(M_B-(1-y)M)-\boldsymbol{r}_T^2+\frac{1}{4}(1-y)^2\boldsymbol{\Delta}_T^2-[M_B-(1-y)M]^2\right]}{D_1(y,\boldsymbol{r}_T,\boldsymbol{\Delta}_T) D_2(\frac{x}{y},\boldsymbol{k}_T-\frac{x}{y}\boldsymbol{r}_T,\boldsymbol{\Delta}_T)}\cdot\frac{(\boldsymbol{k}_T-\frac{x}{y}\boldsymbol{r}_T)\times\boldsymbol{\Delta}_T}{\boldsymbol{k}_T\times\boldsymbol{\Delta}_T},\\
G_{14}\left(x, \boldsymbol{\Delta}_{T}, \boldsymbol{k}_T\right)&=-\frac{g_1^2g_2^2}{8\pi^3}\int_{x}^{1}\frac{dy}{y}\int d^2\boldsymbol{r}_T
\frac{y(1-y)^3(1-\frac{x}{y})^2\cdot\boldsymbol{r}_T\times\boldsymbol{\Delta}_T\cdot(\boldsymbol{k}_T-\frac{x}{y}\boldsymbol{r}_T)\times\boldsymbol{\Delta}_T}{D_1(y,\boldsymbol{r}_T,\boldsymbol{\Delta}_T) D_2(\frac{x}{y},\boldsymbol{k}_T-\frac{x}{y}\boldsymbol{r}_T,\boldsymbol{\Delta}_T)},
\end{align}
where
\begin{align}
\notag D_1(y,\boldsymbol{r}_T,\boldsymbol{\Delta}_T)&=\left[(\boldsymbol{r}_T
-\frac{1}{2}(1-y)\boldsymbol{\Delta}_T)^2+L_1^2\right]^2 \left[(\boldsymbol{r}_T+\frac{1}{2}(1-y)\boldsymbol{\Delta}_T)^2+L_1^2\right]^2,\\	D_2(\frac{x}{y},\boldsymbol{k}_T-\frac{x}{y}\boldsymbol{r}_T,\boldsymbol{\Delta}_T)&
=\left[[(\boldsymbol{k}_T-
\frac{x}{y}\boldsymbol{r}_T)-\frac{1}{2}(1-\frac{x}{y})
\boldsymbol{\Delta}_T]^2+L_2^2\right]^2	\left[[(\boldsymbol{k}_T-\frac{x}{y}\boldsymbol{r}_T)+\frac{1}{2}
(1-\frac{x}{y})\boldsymbol{\Delta}_T]^2+L_2^2\right]^2.
\end{align}
\end{widetext}

Similarly, the Wigner distributions for unpolarized/longitudinally polarized sea quarks in an unpolarized/longitudinally polarized proton can be also calculated from the proton LCWFs within the overlap representation
\begin{widetext} 
\begin{align}\label{UU}
\notag	\rho_{UU}&=\frac{1}{2}\int\frac{d^2\boldsymbol{r}_T}{16\pi^{3}}\int_{x}^{1}\frac{dy}{y}\int \frac{\mathrm{d}^{2} \boldsymbol{\Delta}_T}{(2 \pi)^{2}} e^{-i \boldsymbol{\Delta}_T \cdot \boldsymbol{b}_T} \sum_{\{\lambda\}}\left(\psi_{\lambda_{B}\lambda_{q}\lambda_{\bar{q}}}^{+\star}
\left(x^{out}, y^{out}, \boldsymbol{k}_T^{out},\boldsymbol{r}_T^{out}\right) \psi_{\lambda_{B}\lambda_{q}\lambda_{\bar{q}}}^+\left(x^{in}, y^{in}, \boldsymbol{k}_T^{in},\boldsymbol{r}_T^{in}\right)
	\right.\\&\left.+\psi_{\lambda_{B}\lambda_{q}\lambda_{\bar{q}}}^{-\star}\left(x^{out}, y^{out}, \boldsymbol{k}_T^{out},\boldsymbol{r}_T^{out}\right) \psi_{\lambda_{B}\lambda_{q}\lambda_{\bar{q}}}^-\left(x^{in}, y^{in}, \boldsymbol{k}_T^{in},\boldsymbol{r}_T^{in}\right)\right),\\
\label{LU}	\notag	\rho_{LU}&=\frac{1}{2}\int\frac{d^2\boldsymbol{r}_T}{16\pi^{3}}\int_{x}^{1}\frac{dy}{y}\int \frac{\mathrm{d}^{2} \boldsymbol{\Delta}_T}{(2 \pi)^{2}} e^{-i \boldsymbol{\Delta}_T \cdot \boldsymbol{b}_T} \sum_{\{\lambda\}}\left(\psi_{\lambda_{B}\lambda_{q}\lambda_{\bar{q}}}^{+\star}
\left(x^{out}, y^{out}, \boldsymbol{k}_T^{out},\boldsymbol{r}_T^{out}\right) \psi_{\lambda_{B}\lambda_{q}\lambda_{\bar{q}}}^+\left(x^{in}, y^{in}, \boldsymbol{k}_T^{in},\boldsymbol{r}_T^{in}\right)
	\right.\\&\left.-\psi_{\lambda_{B}\lambda_{q}\lambda_{\bar{q}}}^{-\star}\left(x^{out}, y^{out}, \boldsymbol{k}_T^{out},\boldsymbol{r}_T^{out}\right) \psi_{\lambda_{B}\lambda_{q}\lambda_{\bar{q}}}^-\left(x^{in}, y^{in}, \boldsymbol{k}_T^{in},\boldsymbol{r}_T^{in}\right)\right),\\
\label{UL}
	\notag \rho_{UL}&=\frac{1}{2}\int\frac{d^2\boldsymbol{r}_T}{16\pi^{3}}\int_{x}^{1}\frac{dy}{y}\int \frac{\mathrm{d}^{2} \boldsymbol{\Delta}_T}{(2 \pi)^{2}} e^{-i \boldsymbol{\Delta}_T \cdot \boldsymbol{b}_T} sign(\lambda_{\bar{q}})\sum_{\{\lambda\}}\left(\psi_{\lambda_{B}\lambda_{q}
\lambda_{\bar{q}}}^{+\star}\left(x^{out}, y^{out}, \boldsymbol{k}_T^{out},\boldsymbol{r}_T^{out}\right) \psi_{\lambda_{B}\lambda_{q}\lambda_{\bar{q}}}^+\left(x^{in}, y^{in}, \boldsymbol{k}_T^{in},\boldsymbol{r}_T^{in}\right)
	\right.\\&\left.+\psi_{\lambda_{B}\lambda_{q}\lambda_{\bar{q}}}^{-\star}\left(x^{out}, y^{out}, \boldsymbol{k}_T^{out},\boldsymbol{r}_T^{out}\right) \psi_{\lambda_{B}\lambda_{q}\lambda_{\bar{q}}}^-\left(x^{in}, y^{in}, \boldsymbol{k}_T^{in},\boldsymbol{r}_T^{in}\right)\right),
\\
\label{LL}\notag \rho_{LL}&=\frac{1}{2}\int\frac{d^2\boldsymbol{r}_T}{16\pi^{3}}\int_{x}^{1}\frac{dy}{y}\int \frac{\mathrm{d}^{2} \boldsymbol{\Delta}_T}{(2 \pi)^{2}} e^{-i \boldsymbol{\Delta}_T \cdot \boldsymbol{b}_T} sign(\lambda_{\bar{q}})\sum_{\{\lambda\}}\left(\psi_{\lambda_{B}\lambda_{q}
\lambda_{\bar{q}}}^{+\star}\left(x^{out}, y^{out}, \boldsymbol{k}_T^{out},\boldsymbol{r}_T^{out}\right) \psi_{\lambda_{B}\lambda_{q}\lambda_{\bar{q}}}^+\left(x^{in}, y^{in}, \boldsymbol{k}_T^{in},\boldsymbol{r}_T^{in}\right)
	\right.\\&\left.-\psi_{\lambda_{B}\lambda_{q}\lambda_{\bar{q}}}^{-\star}\left(x^{out}, y^{out}, \boldsymbol{k}_T^{out},\boldsymbol{r}_T^{out}\right) \psi_{\lambda_{B}\lambda_{q}\lambda_{\bar{q}}}^-\left(x^{in}, y^{in}, \boldsymbol{k}_T^{in},\boldsymbol{r}_T^{in}\right)\right).
\end{align}
\end{widetext}
Substituting the LCWFs of the proton into Eqs.~(\ref{UU}), (\ref{LU}), (\ref{UL}) and (\ref{LL}), we obtain the expressions for the Wigner distributions in our model as follows:
\begin{widetext}
\begin{align}
\notag
		\rho_{UU}\left(x, \boldsymbol{b}_T, \boldsymbol{k}_T\right)&=\frac{g_1^2g_2^2}{(2\pi)^5}\int_{x}^{1}\frac{dy}{y}\int d^2\boldsymbol{r}_T\int d^2 \boldsymbol{\Delta}_T e^{-i \boldsymbol{\Delta}_T \cdot \boldsymbol{b}_T} 		\\&\frac{y(1-y)^2(1-\frac{x}{y})^2\left[\boldsymbol{r}_T^2-\frac{1}{4}(1-y)^2
\boldsymbol{\Delta}_T^2+[M_B-(1-y)M]^2\right]\left[(\boldsymbol{k}_T-\frac{x}{y}
\boldsymbol{r}_T)^2-\frac{1}{4}(1-\frac{x}{y})^2\boldsymbol{\Delta}_T^2+m^2\right]}
{D_1(y,\boldsymbol{r}_T,\boldsymbol{\Delta}_T) D_2(\frac{x}{y},\boldsymbol{k}_T-\frac{x}{y}\boldsymbol{r}_T,\boldsymbol{\Delta}_T)},\\
		\notag
		\rho_{LU}\left(x, \boldsymbol{b}_T, \boldsymbol{k}_T\right)&=\frac{ig_1^2g_2^2}{(2\pi)^5}\int_{x}^{1}\frac{dy}{y}\int d^2\boldsymbol{r}_T\int d^2 \boldsymbol{\Delta}_T e^{-i \boldsymbol{\Delta}_T \cdot \boldsymbol{b}_T} 		\\&\frac{y(1-y)^3(1-\frac{x}{y})^2\left[(\boldsymbol{k}_T-\frac{x}{y}
\boldsymbol{r}_T)^2-\frac{1}{4}(1-\frac{x}{y})^2\boldsymbol{\Delta}_T^2+m^2\right]
\cdot\boldsymbol{r}_T\times\boldsymbol{\Delta}_T}
{D_1(y,\boldsymbol{r}_T,\boldsymbol{\Delta}_T) D_2(\frac{x}{y},\boldsymbol{k}_T-\frac{x}{y}\boldsymbol{r}_T,\boldsymbol{\Delta}_T)},\\
		\notag
		\rho_{UL}\left(x, \boldsymbol{b}_T, \boldsymbol{k}_T\right)&=\frac{ig_1^2g_2^2}{(2\pi)^5}\int_{x}^{1}\frac{dy}{y}\int d^2\boldsymbol{r}_T\int d^2 \boldsymbol{\Delta}_T e^{-i \boldsymbol{\Delta}_T \cdot \boldsymbol{b}_T} \\&\frac{y(1-y)^2(1-\frac{x}{y})^3\left[\boldsymbol{r}_T^2-\frac{1}{4}(1-y)^2
\boldsymbol{\Delta}_T^2+[M_B-(1-y)M]^2\right]\cdot(\boldsymbol{k}_T-\frac{x}{y}
\boldsymbol{r}_T)\times\boldsymbol{\Delta}_T}{D_1(y,\boldsymbol{r}_T,\boldsymbol{\Delta}_T) D_2(\frac{x}{y},\boldsymbol{k}_T-\frac{x}{y}\boldsymbol{r}_T,\boldsymbol{\Delta}_T)},\\
		\rho_{LL}\left(x, \boldsymbol{b}_T, \boldsymbol{k}_T\right)&=-\frac{g_1^2g_2^2}{(2\pi)^5}\int_{x}^{1}\frac{dy}{y}\int d^2\boldsymbol{r}_T\int d^2 \boldsymbol{\Delta}_T e^{-i \boldsymbol{\Delta}_T \cdot \boldsymbol{b}_T} 	\frac{y(1-y)^3(1-\frac{x}{y})^2\cdot\boldsymbol{r}_T\times\boldsymbol{\Delta}_T\cdot(\boldsymbol{k}_T-\frac{x}{y}\boldsymbol{r}_T)\times\boldsymbol{\Delta}_T}{D_1(y,\boldsymbol{r}_T,\boldsymbol{\Delta}_T) D_2(\frac{x}{y},\boldsymbol{k}_T-\frac{x}{y}\boldsymbol{r}_T,\boldsymbol{\Delta}_T)}.
	\end{align}
\end{widetext}

\section{Numerical results for sea quark Wigner distributions }\label{Sec3}

\begin{figure*}[htbp]
		\centering
		\subfigure{\begin{minipage}[b]{0.4\linewidth}
				\centering
				\includegraphics[width=\linewidth]{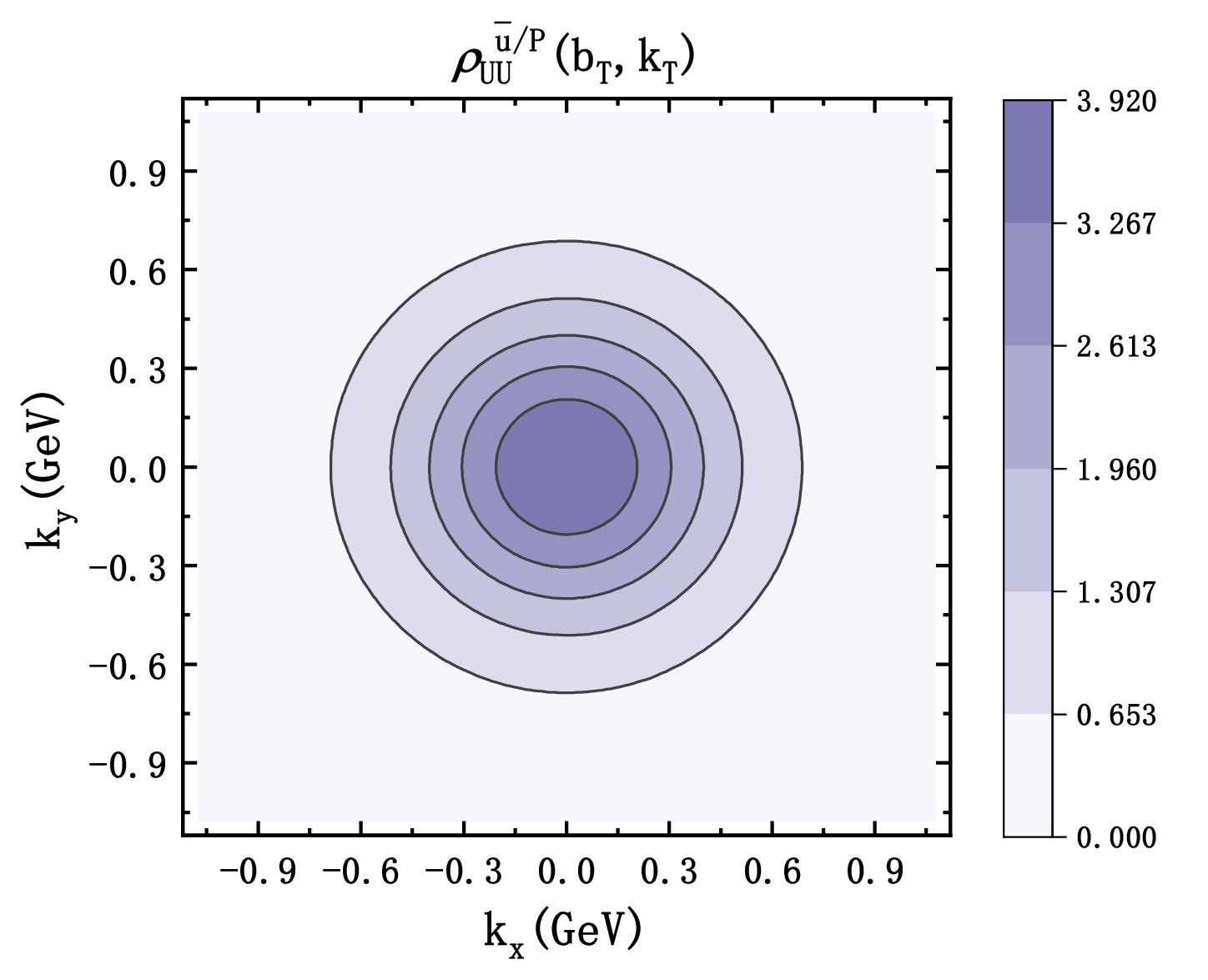}
		\end{minipage}}
		\subfigure{\begin{minipage}[b]{0.4\linewidth}
				\centering
				\includegraphics[width=\linewidth]{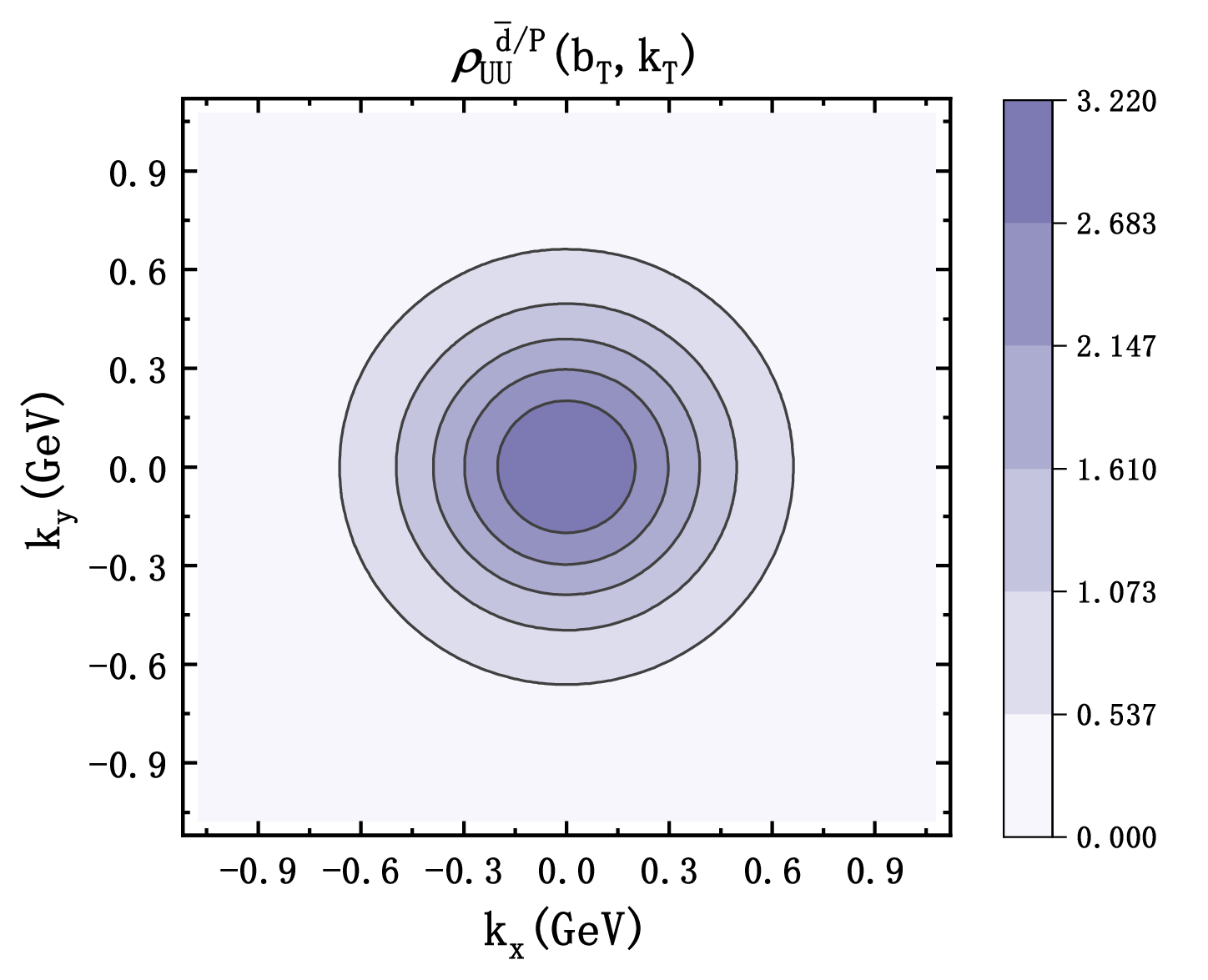}    	
\end{minipage}}
		\subfigure{\begin{minipage}[b]{0.4\linewidth}
				\centering
				\includegraphics[width=\linewidth]{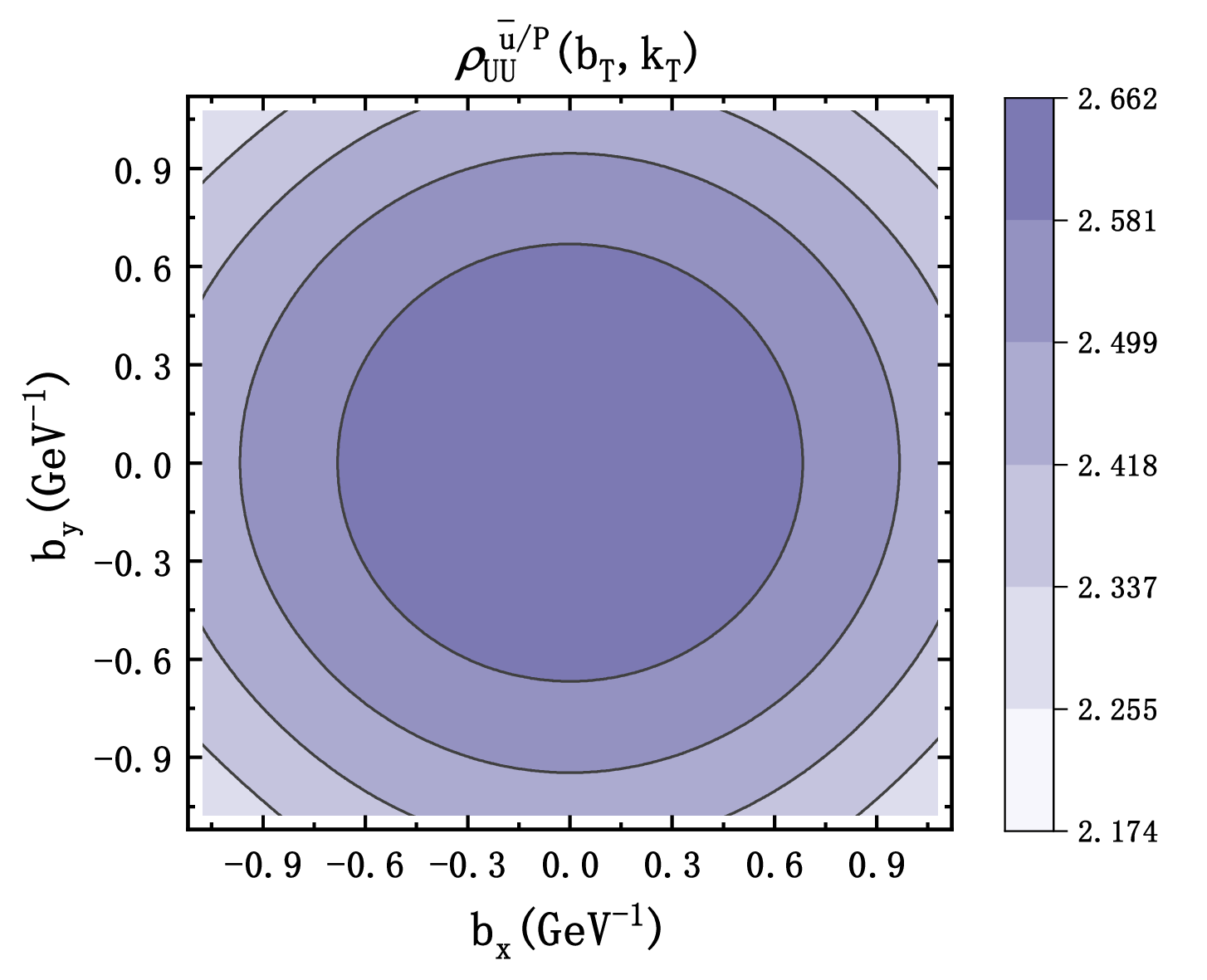}
		\end{minipage}}
		\subfigure{\begin{minipage}[b]{0.4\linewidth}
				\centering
				\includegraphics[width=\linewidth]{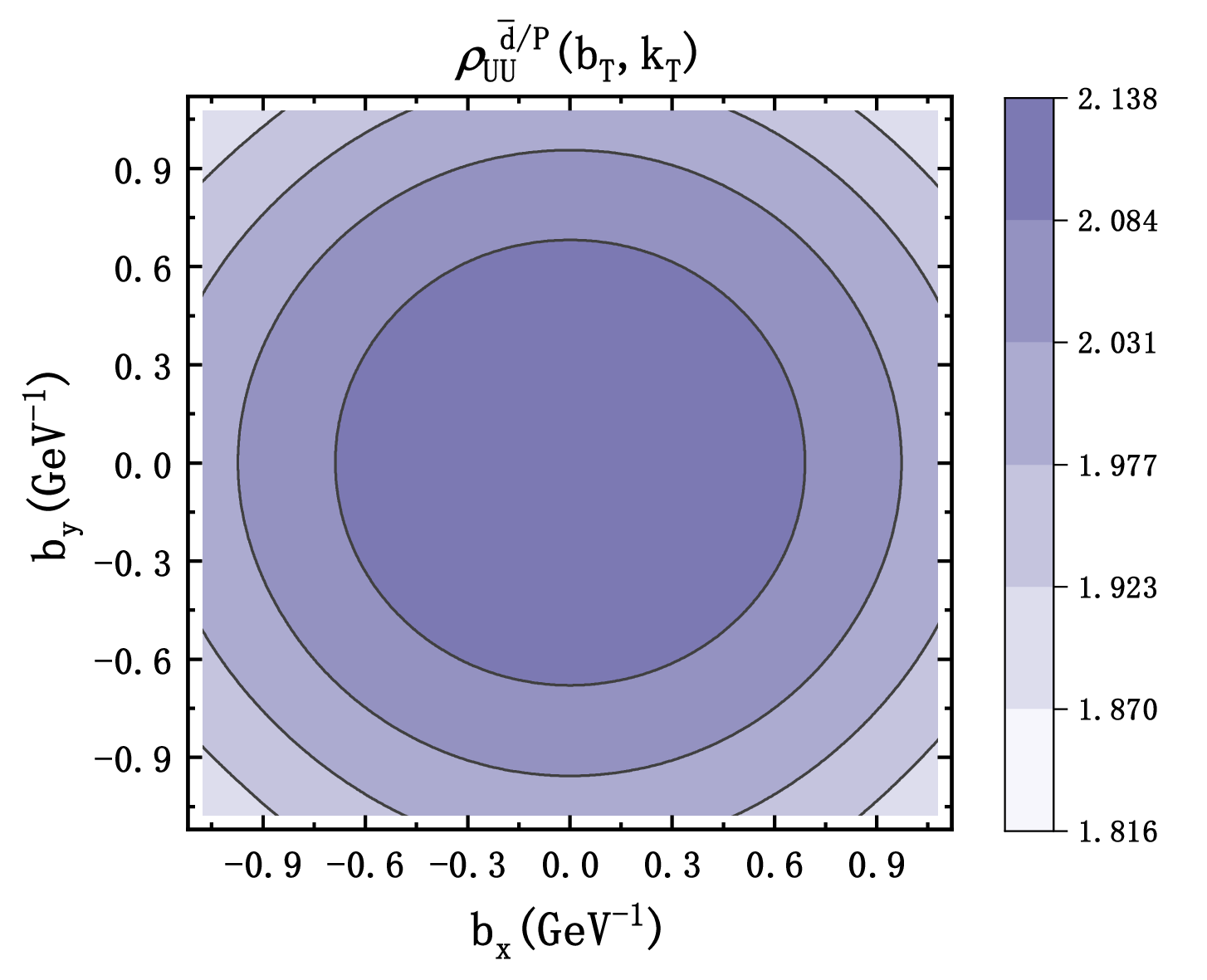}  
\end{minipage}}
		\subfigure{\begin{minipage}[b]{0.4\linewidth}
				\centering
				\includegraphics[width=\linewidth]{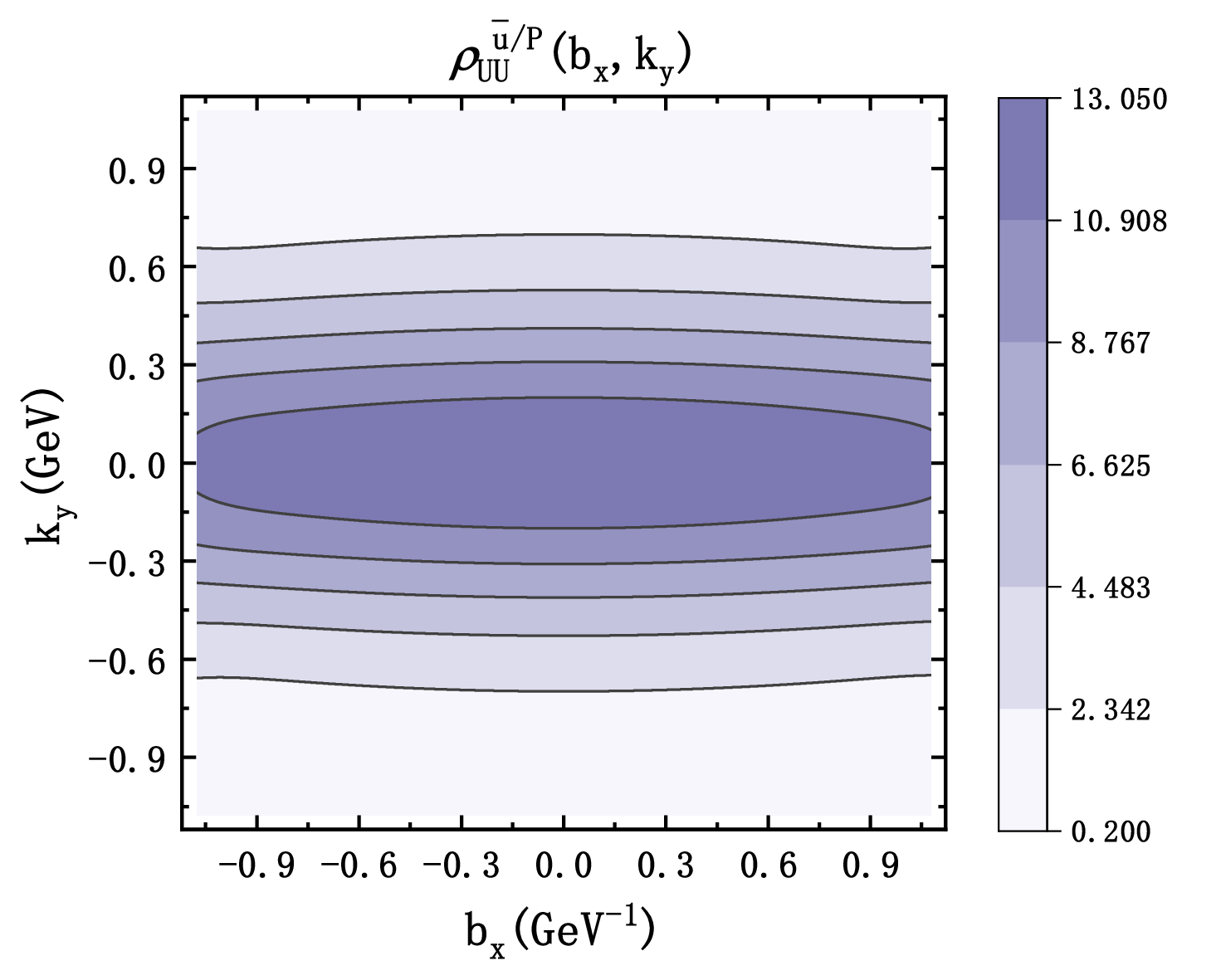}
		\end{minipage}}
		\subfigure{\begin{minipage}[b]{0.4\linewidth}
				\centering
				\includegraphics[width=\linewidth]{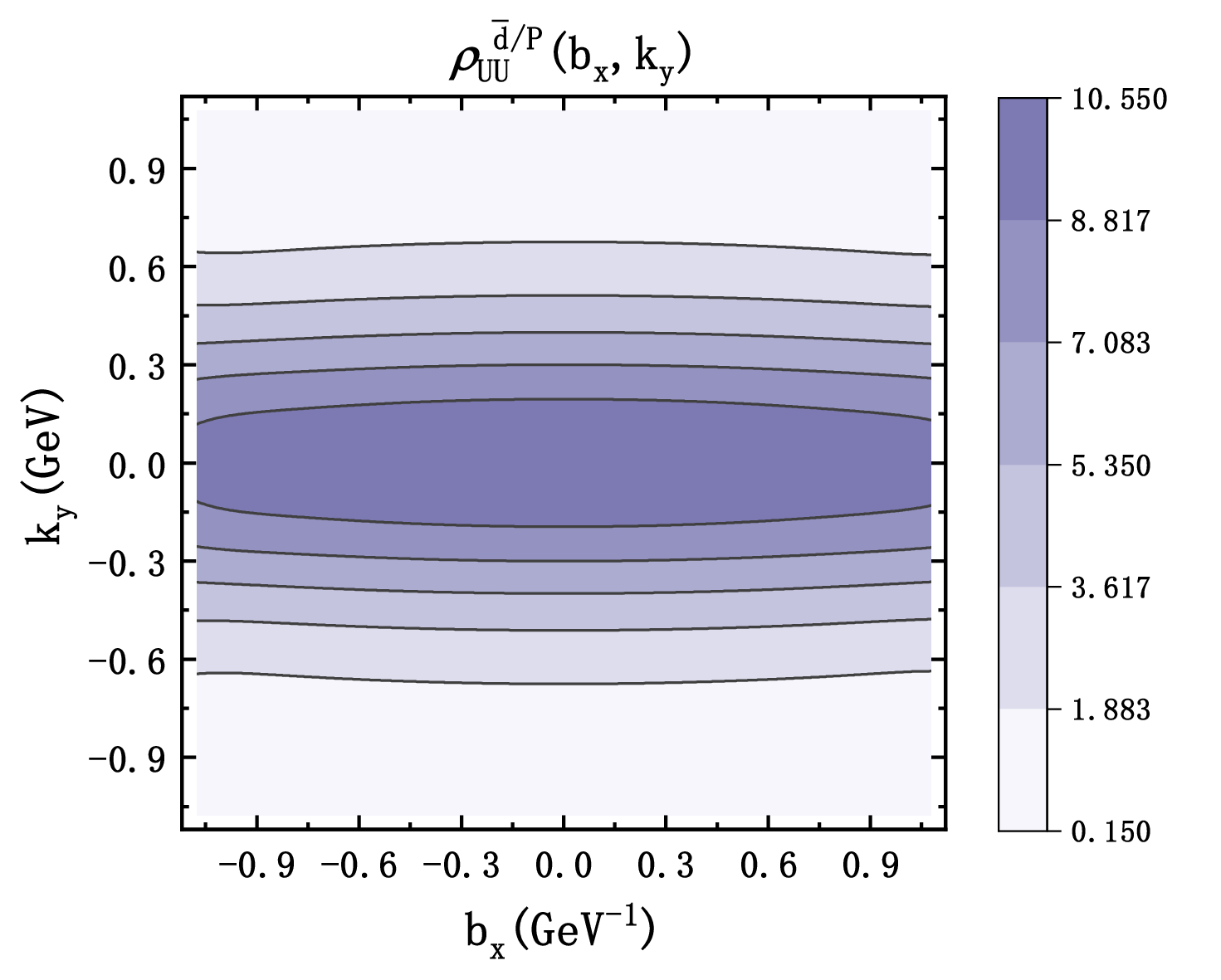}    	
\end{minipage}}
\caption{The Wigner distribution $\rho_{UU}$ of the $\bar{u}$ (left panel) and $\bar{d}$ (right panel) quarks in the proton. 
The upper panel depicts the distributions in the transverse momentum space with fixed impact parameter $\boldsymbol{b}_{T}=0.3$ GeV$^{-1} \ \hat{\boldsymbol{e}}_{y}$. 
The central panel depicts the distributions in the impact parameter space with fixed transverse momentum $\boldsymbol{k}_{T}=0.3$ GeV $\hat{\boldsymbol{e}}_{y}$.
The lower panel depicts the distributions in the mixed plane.} 
\label{uu}      
\end{figure*}
	
\begin{figure*}[htbp]
		\centering
		\subfigure{\begin{minipage}[b]{0.4\linewidth}
				\centering
				\includegraphics[width=\linewidth]{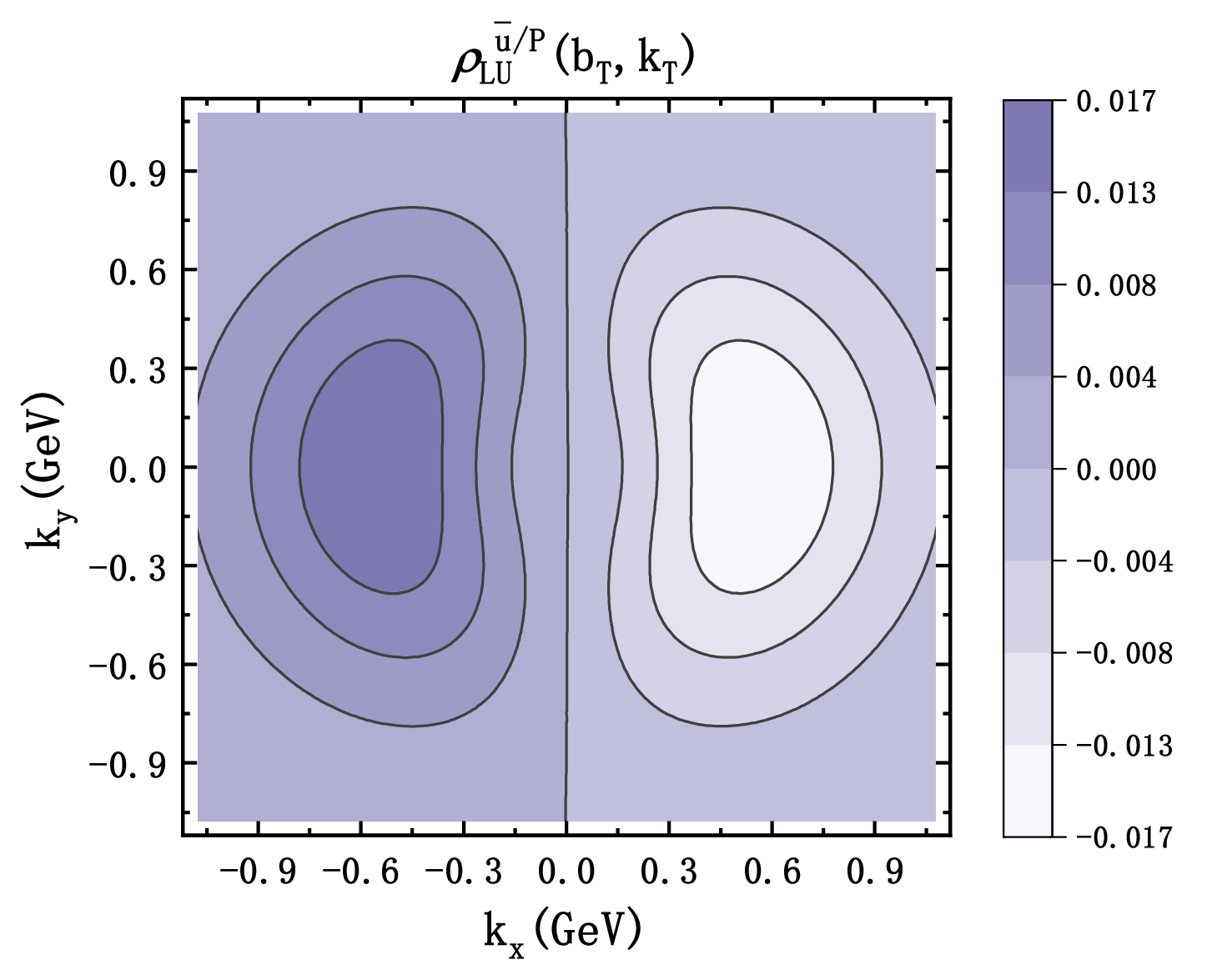}
		\end{minipage}}
		\subfigure{\begin{minipage}[b]{0.4\linewidth}
				\centering
				\includegraphics[width=\linewidth]{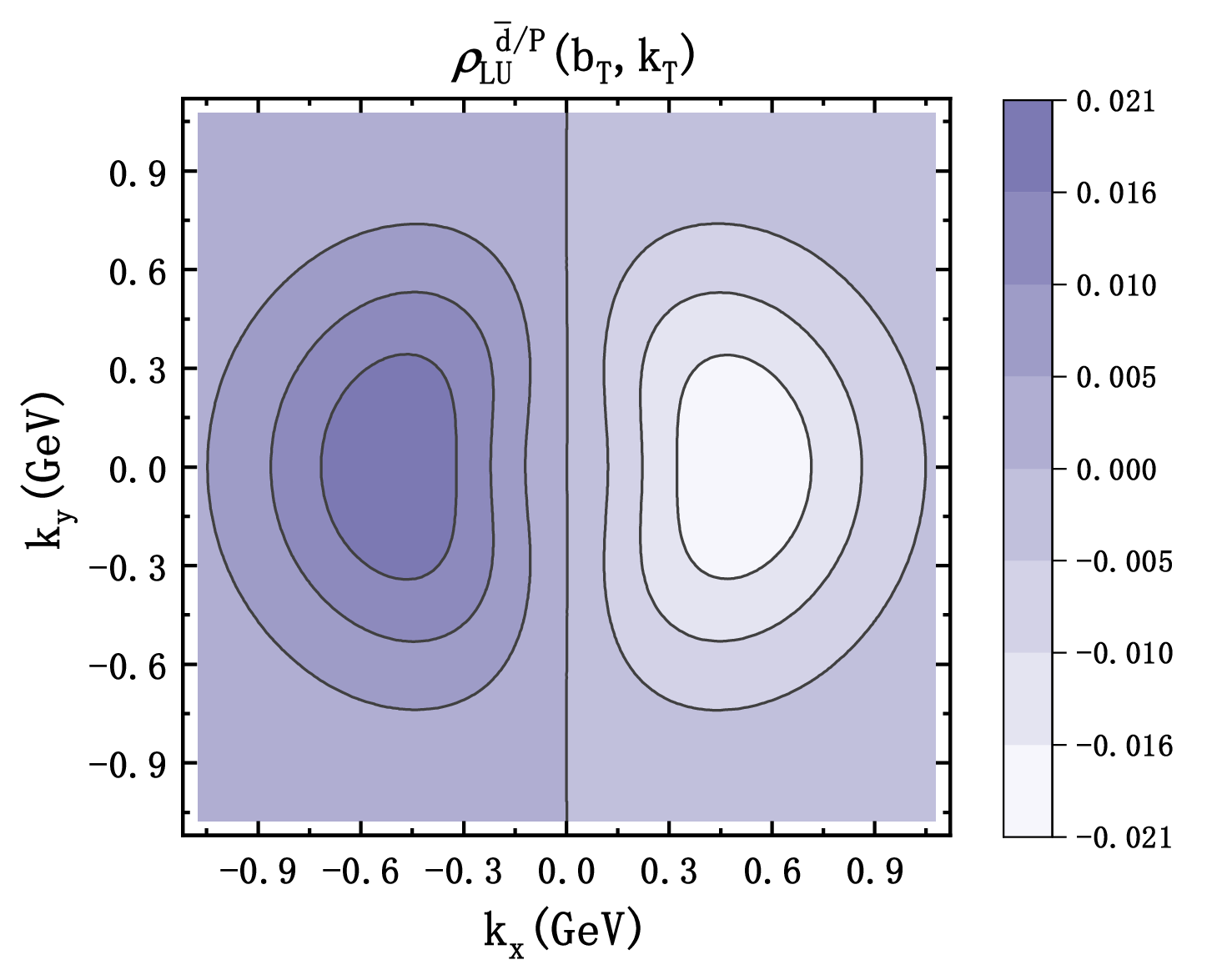}    	
\end{minipage}}
		\subfigure{\begin{minipage}[b]{0.4\linewidth}
				\centering
				\includegraphics[width=\linewidth]{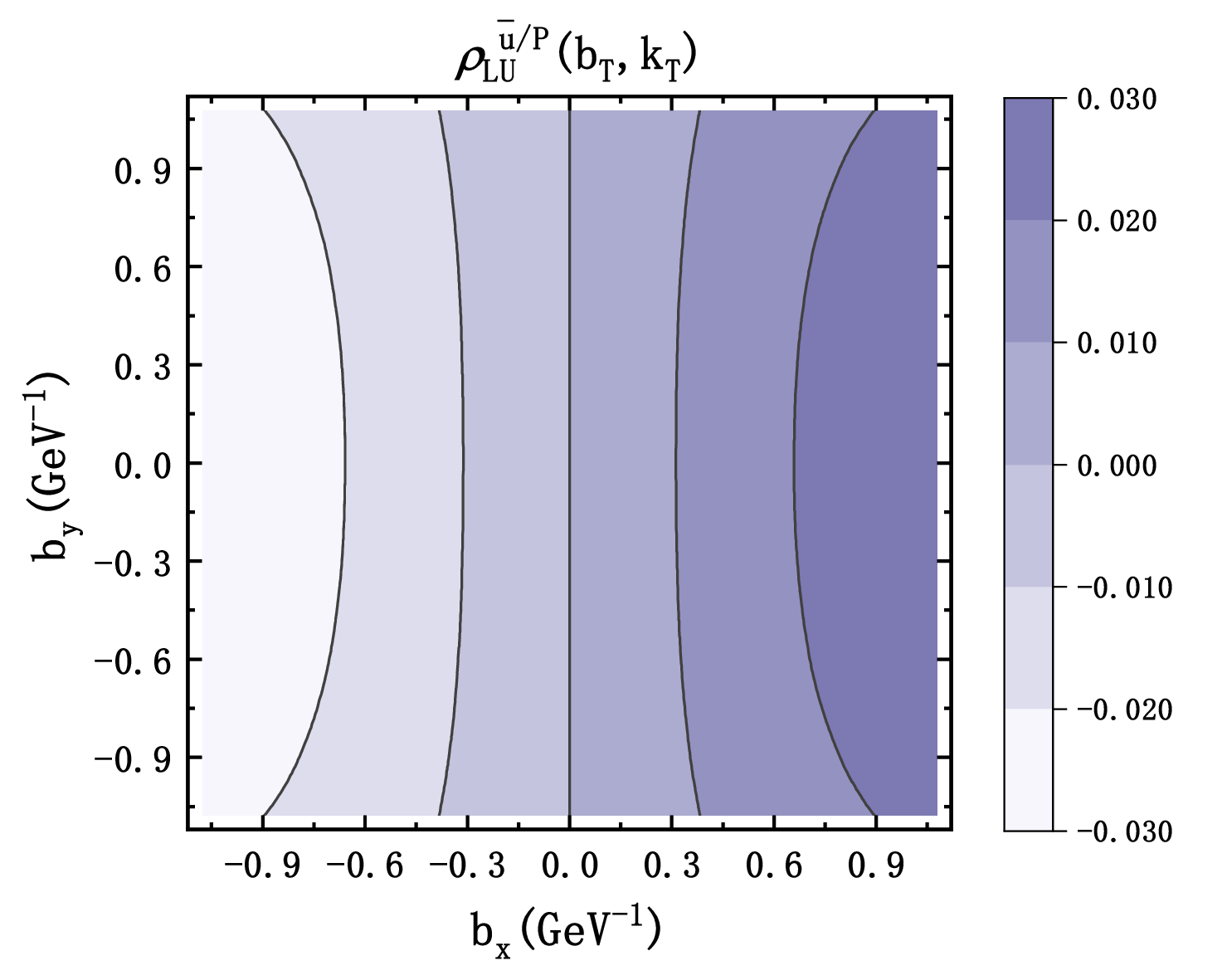}
		\end{minipage}}
		\subfigure{\begin{minipage}[b]{0.4\linewidth}
				\centering
				\includegraphics[width=\linewidth]{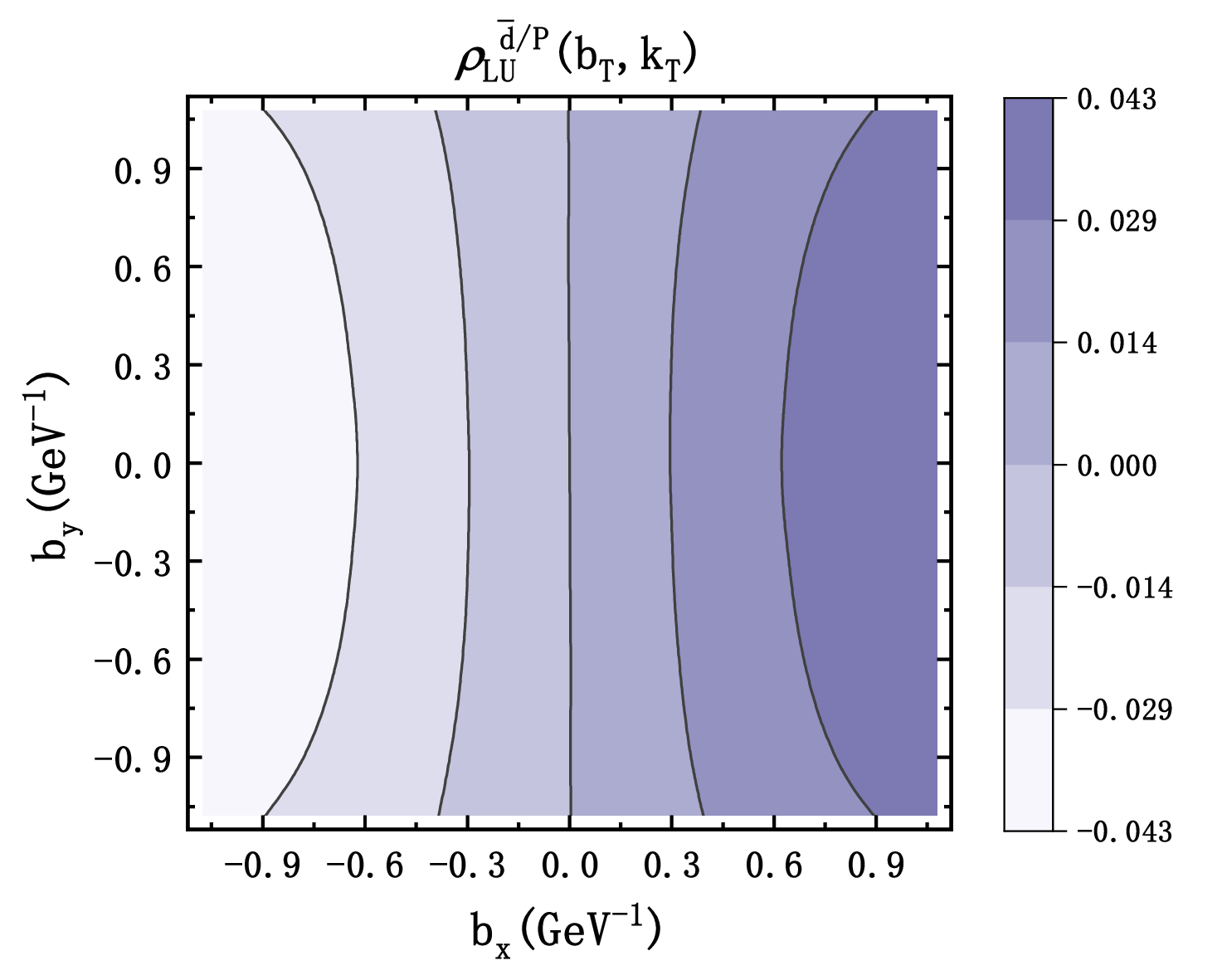}  
\end{minipage}}
		\subfigure{\begin{minipage}[b]{0.4\linewidth}
				\centering
				\includegraphics[width=\linewidth]{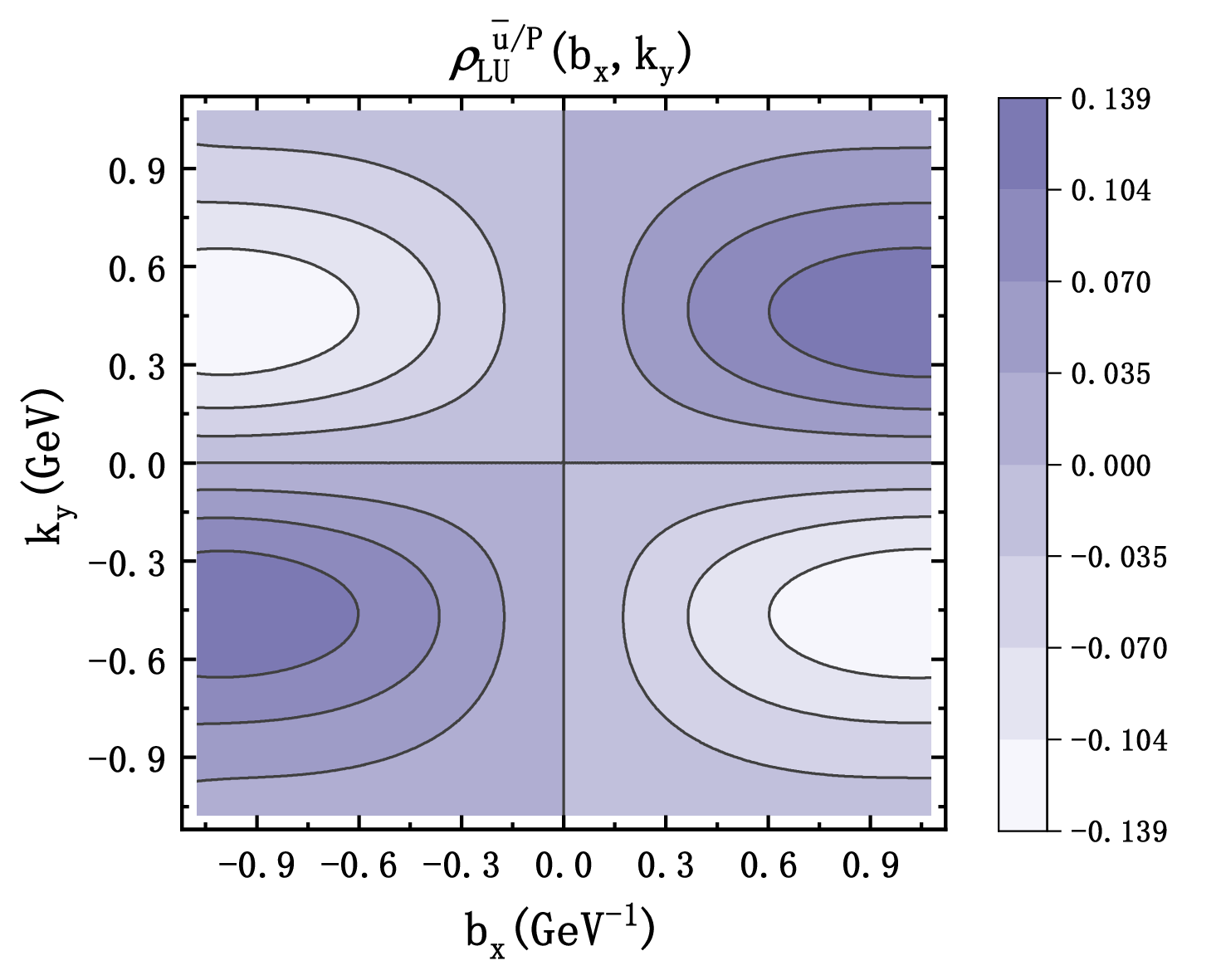}
		\end{minipage}}
		\subfigure{\begin{minipage}[b]{0.4\linewidth}
				\centering
				\includegraphics[width=\linewidth]{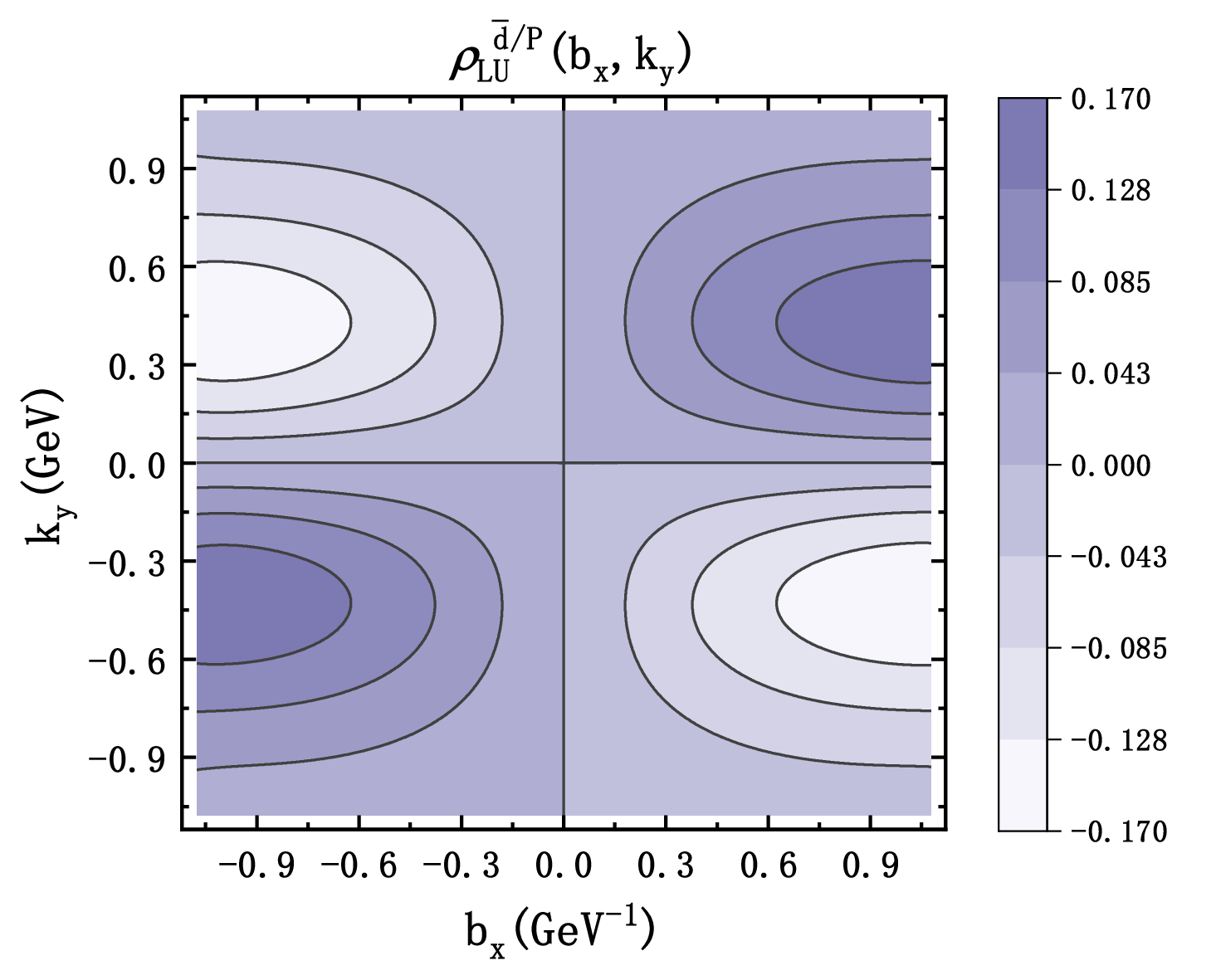}    	
\end{minipage}}
\caption{Similar to Fig.~\ref{uu}, but for the Wigner distribution $\rho_{LU}$ of the $\bar{u}$ (left panel) and $\bar{d}$ (right panel) quarks in the proton.} \label{lu}       
\end{figure*}
	
\begin{figure*}[htbp]
		\centering
		\subfigure{\begin{minipage}[b]{0.4\linewidth}
				\centering
				\includegraphics[width=\linewidth]{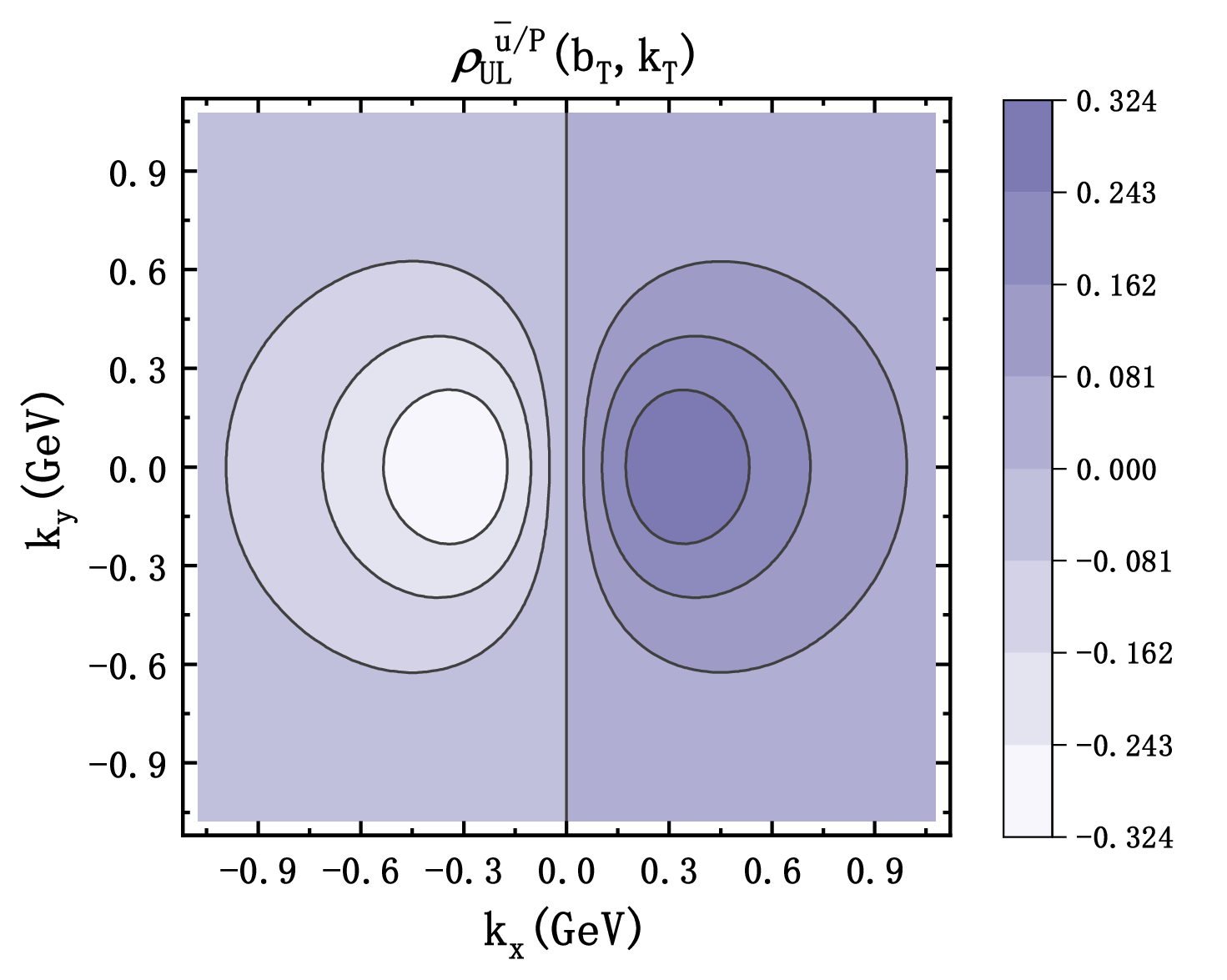}
		\end{minipage}}
		\subfigure{\begin{minipage}[b]{0.4\linewidth}
				\centering
				\includegraphics[width=\linewidth]{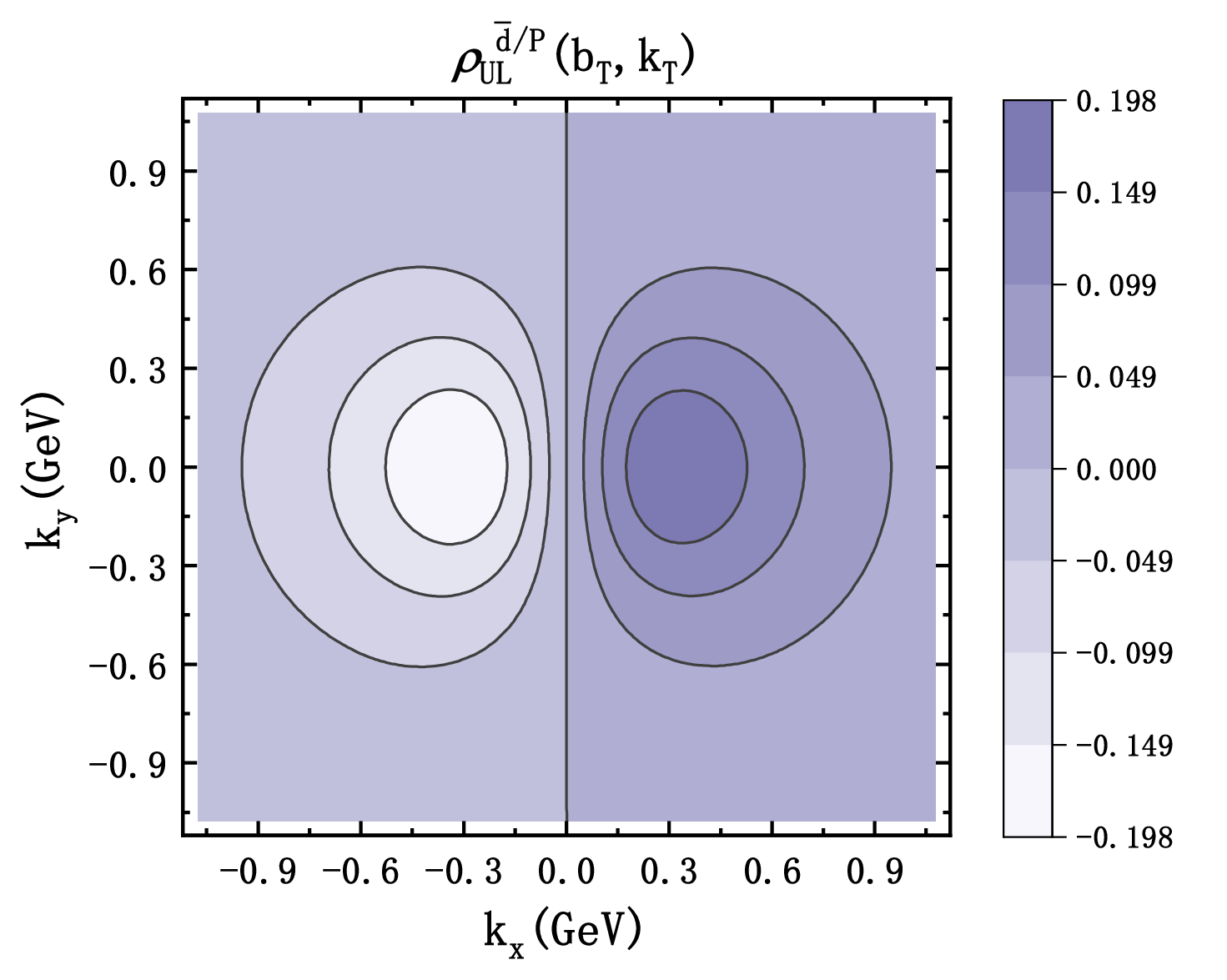}    	\end{minipage}}
		\subfigure{\begin{minipage}[b]{0.4\linewidth}
				\centering
				\includegraphics[width=\linewidth]{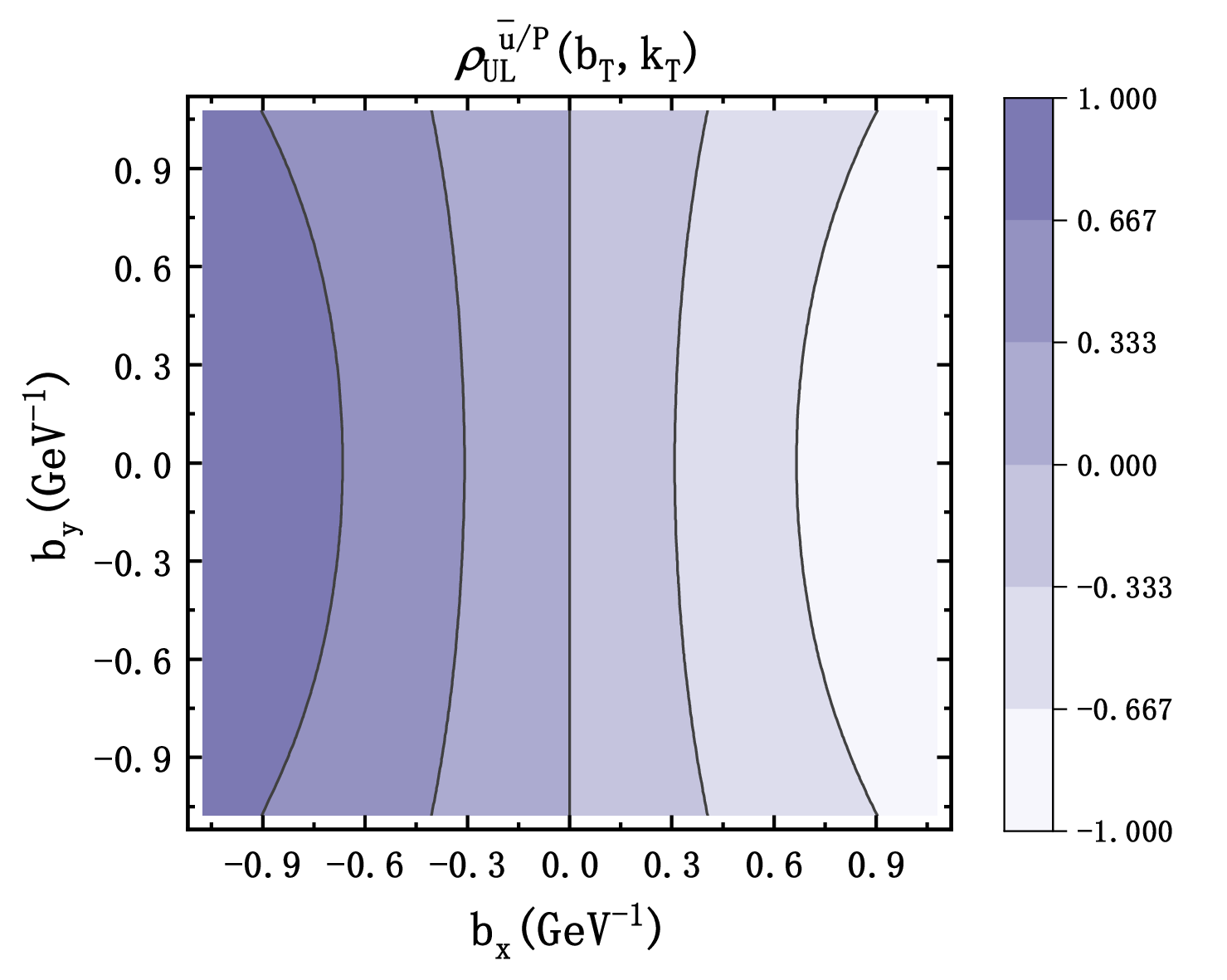}
		\end{minipage}}
		\subfigure{\begin{minipage}[b]{0.4\linewidth}
				\centering
				\includegraphics[width=\linewidth]{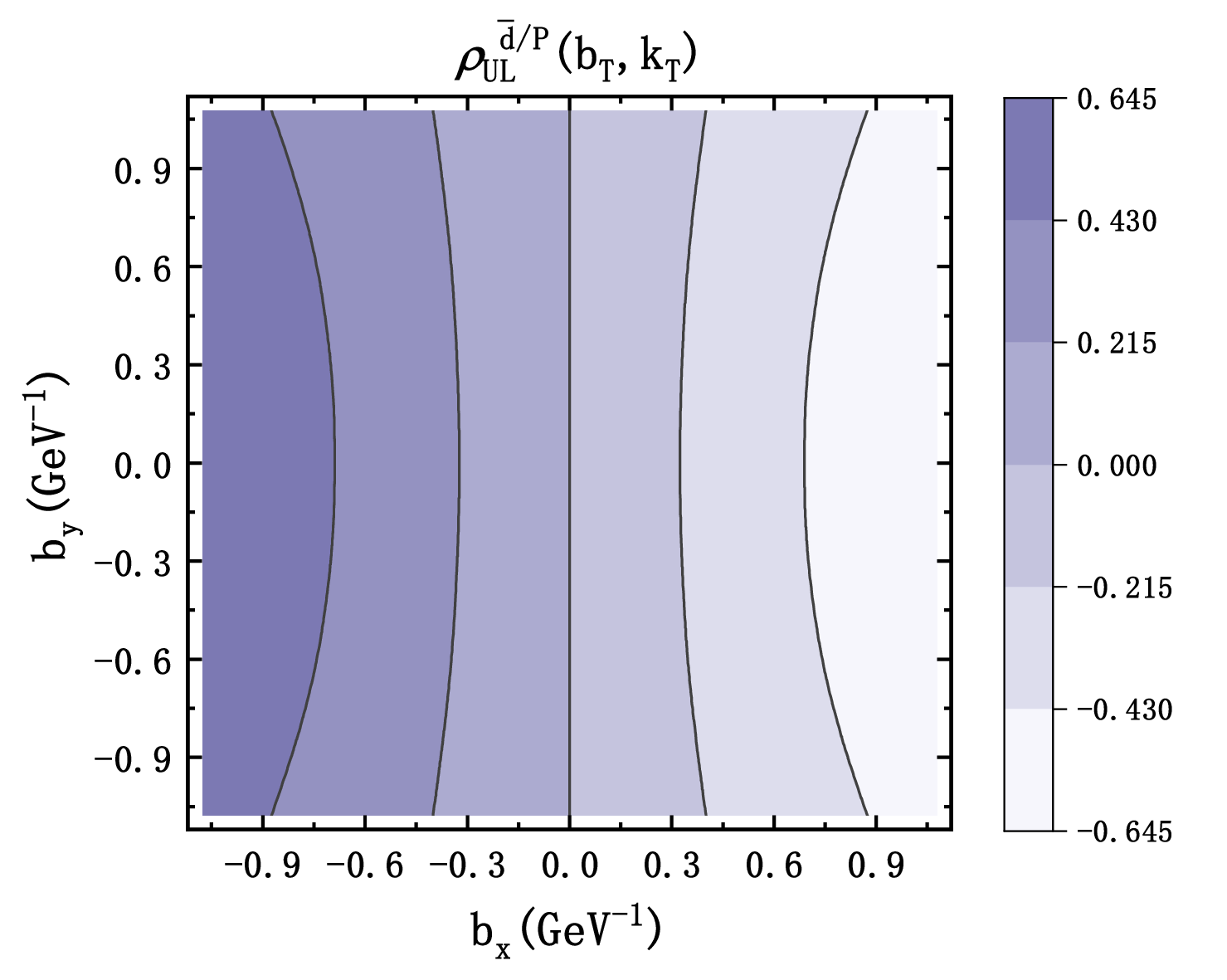}  \end{minipage}}
		\subfigure{\begin{minipage}[b]{0.4\linewidth}
				\centering
				\includegraphics[width=\linewidth]{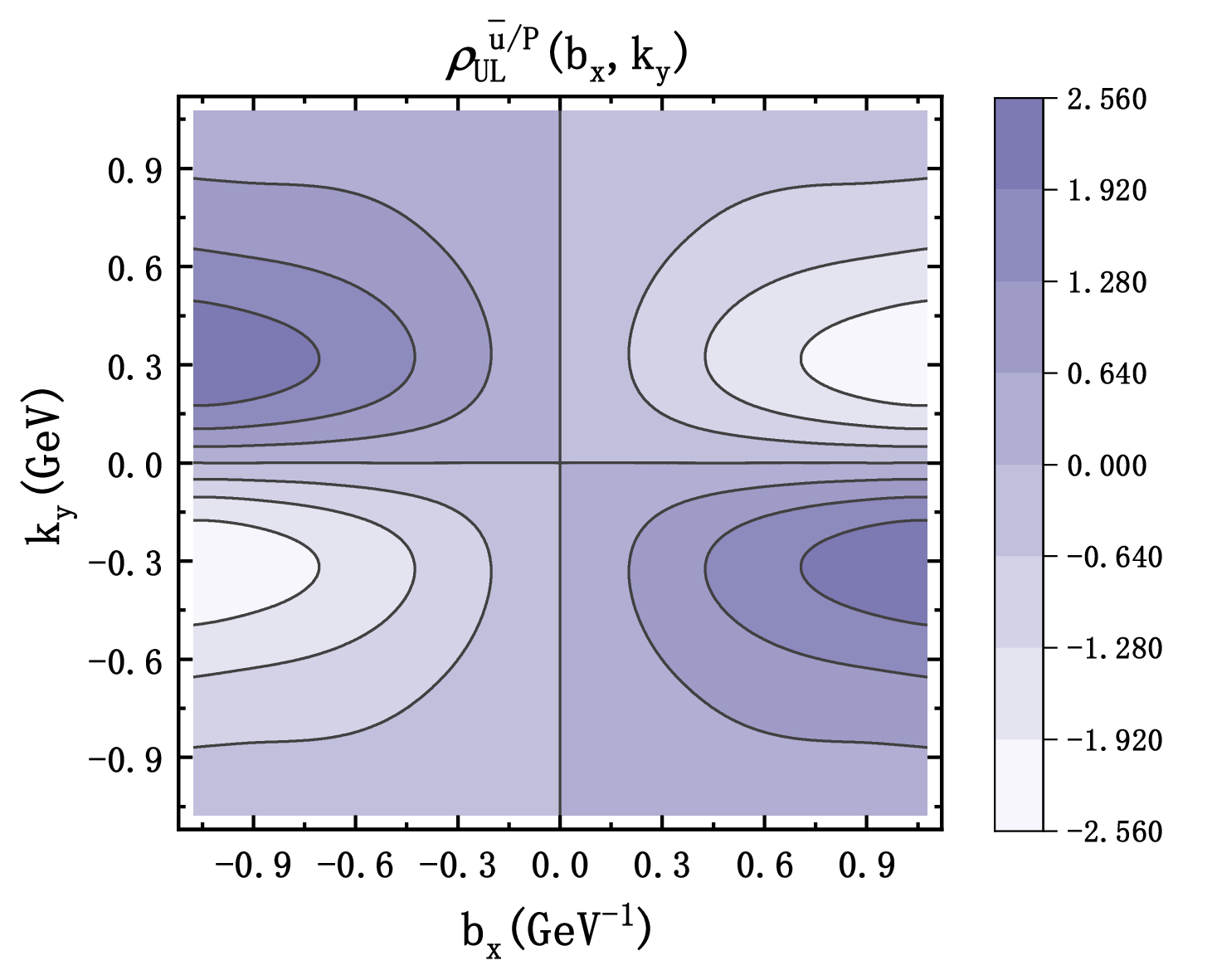}
		\end{minipage}}
		\subfigure{\begin{minipage}[b]{0.4\linewidth}
				\centering
				\includegraphics[width=\linewidth]{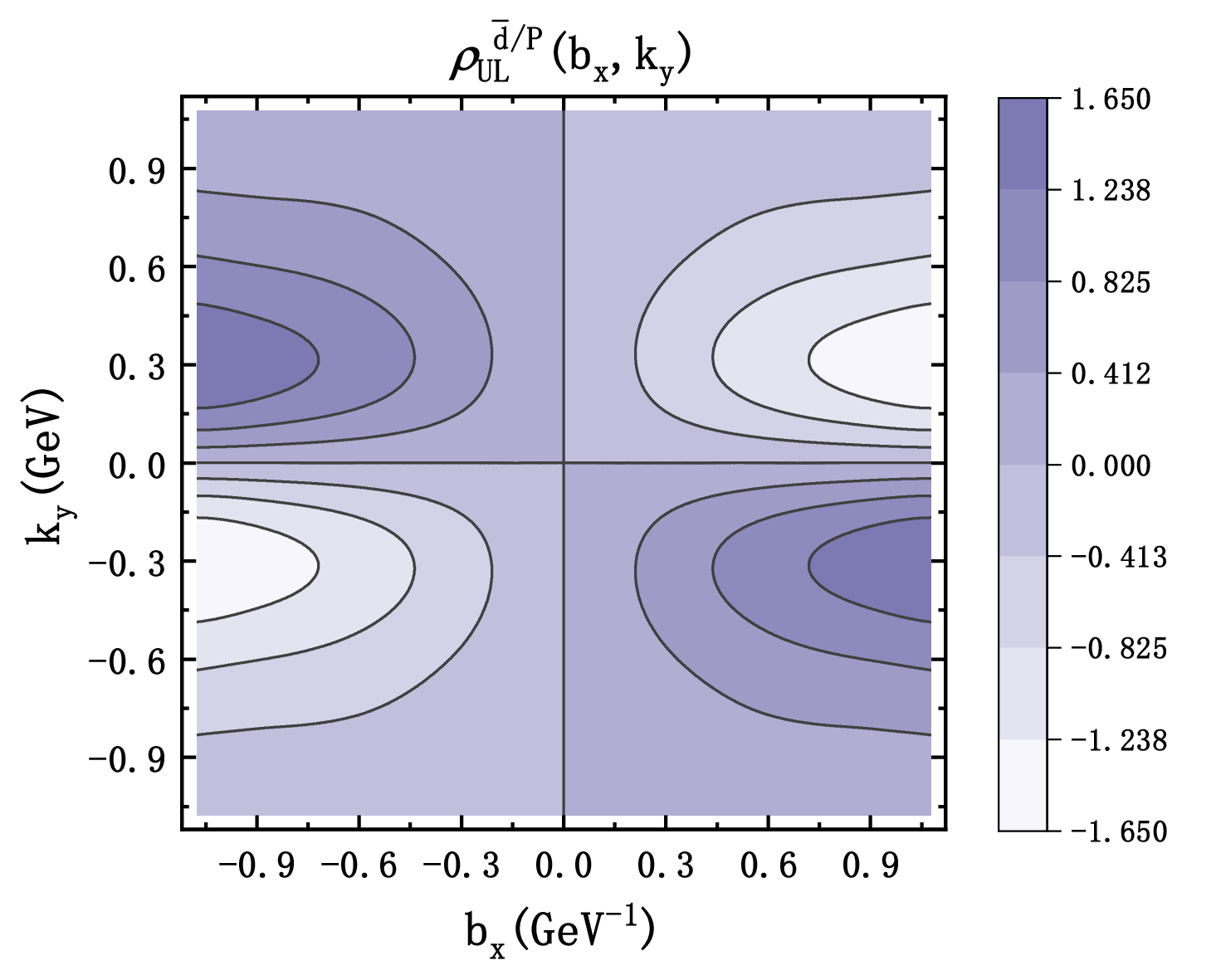}    	\end{minipage}}
\caption{Similar to Fig.~\ref{uu}, but for the Wigner distribution $\rho_{UL}$ of the $\bar{u}$ (left panel) and $\bar{d}$ (right panel) quarks in the proton.} \label{ul}      
	\end{figure*}
	
	\begin{figure*}[htbp]
		\centering
		\subfigure{\begin{minipage}[b]{0.4\linewidth}
				\centering
				\includegraphics[width=\linewidth]{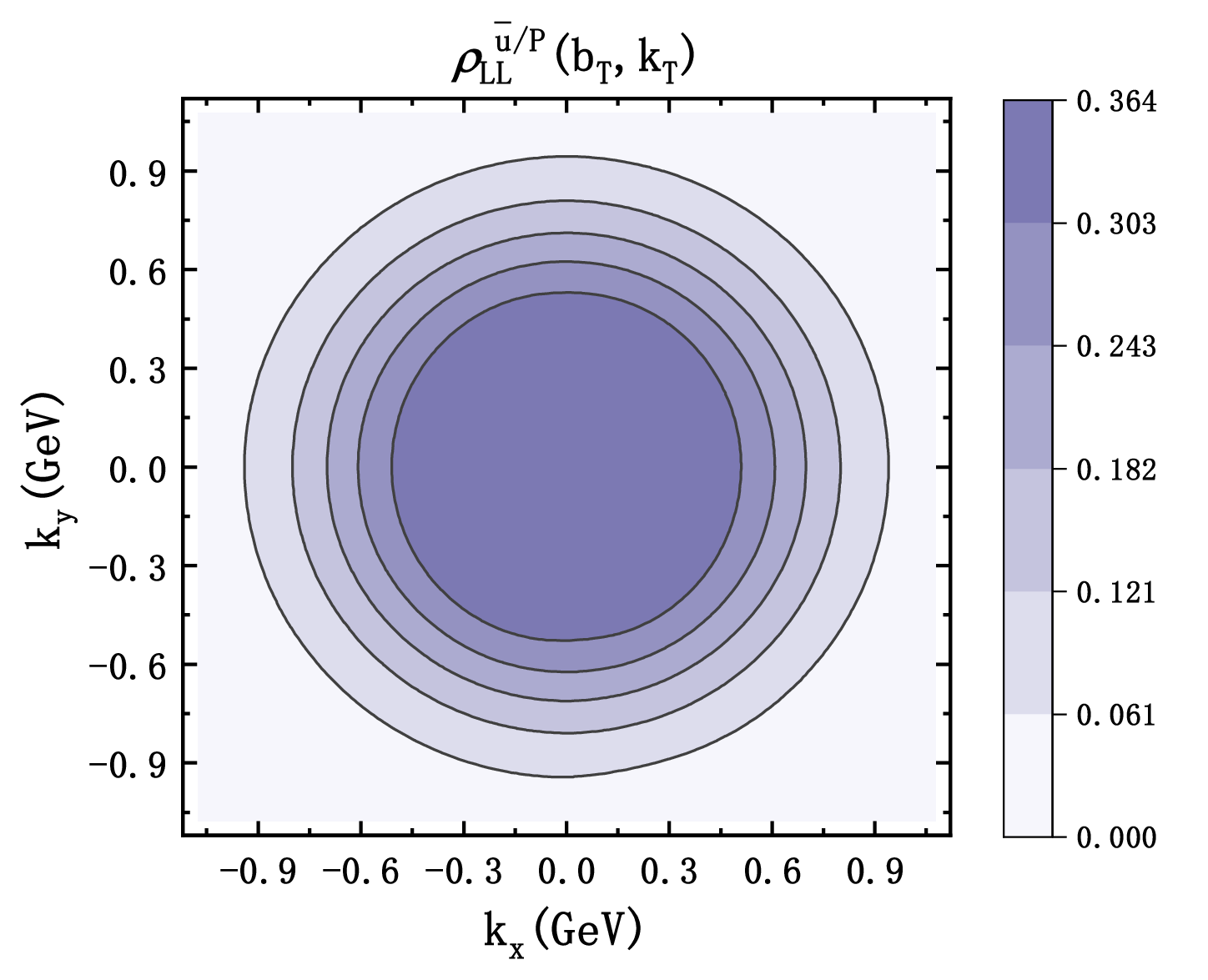}
		\end{minipage}}
		\subfigure{\begin{minipage}[b]{0.4\linewidth}
				\centering
				\includegraphics[width=\linewidth]{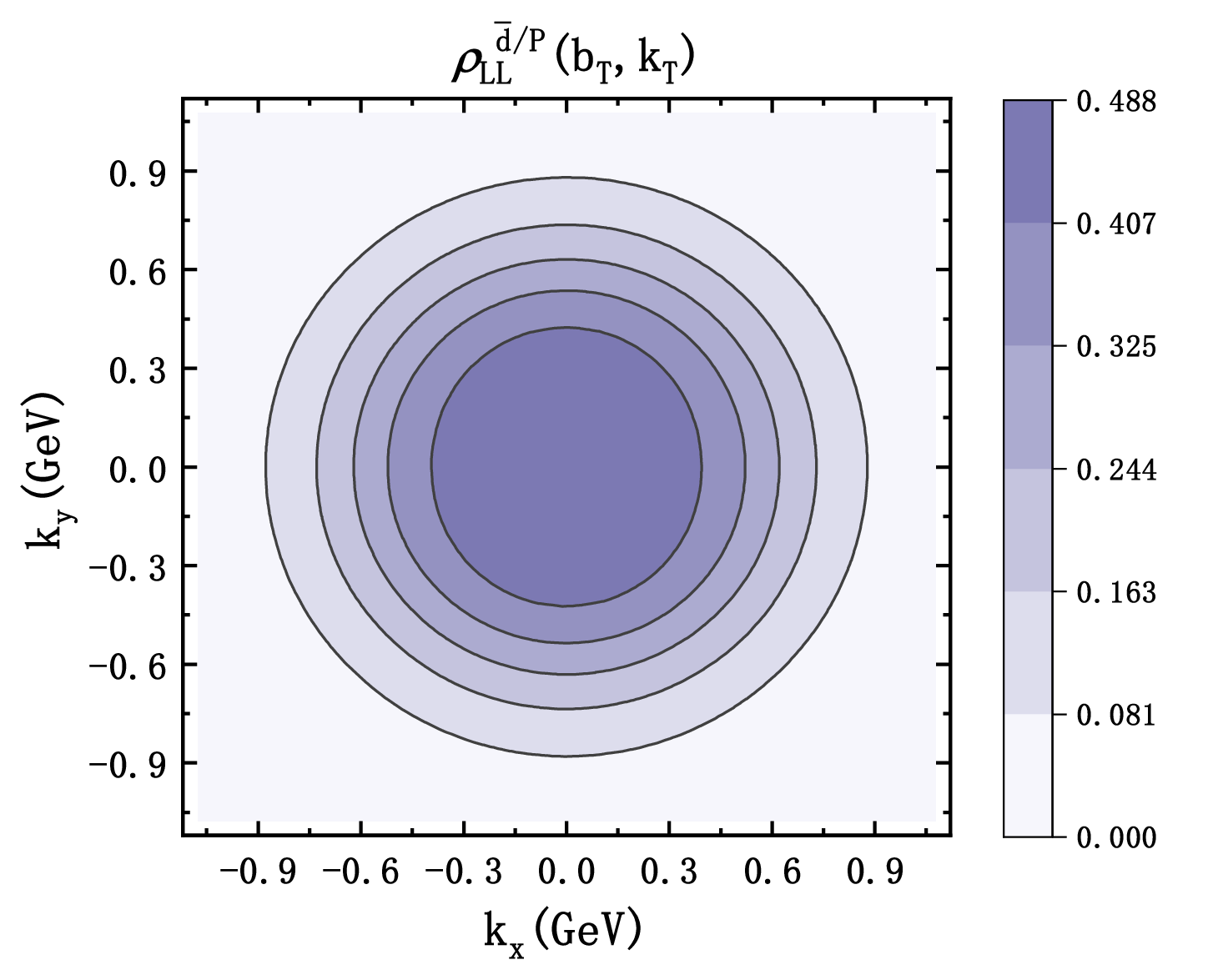}    	\end{minipage}}
		\subfigure{\begin{minipage}[b]{0.4\linewidth}
				\centering
				\includegraphics[width=\linewidth]{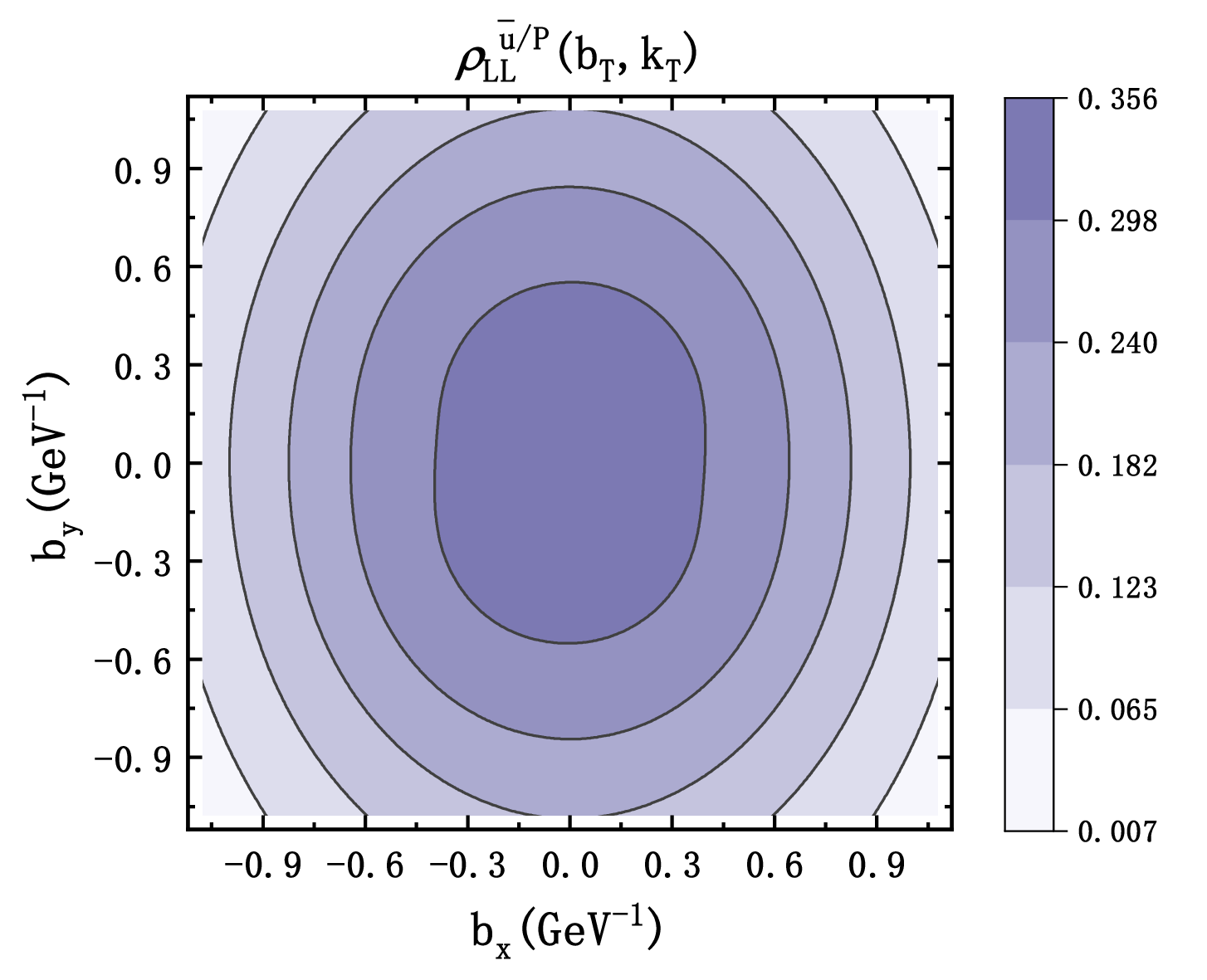}
		\end{minipage}}
		\subfigure{\begin{minipage}[b]{0.4\linewidth}
				\centering
				\includegraphics[width=\linewidth]{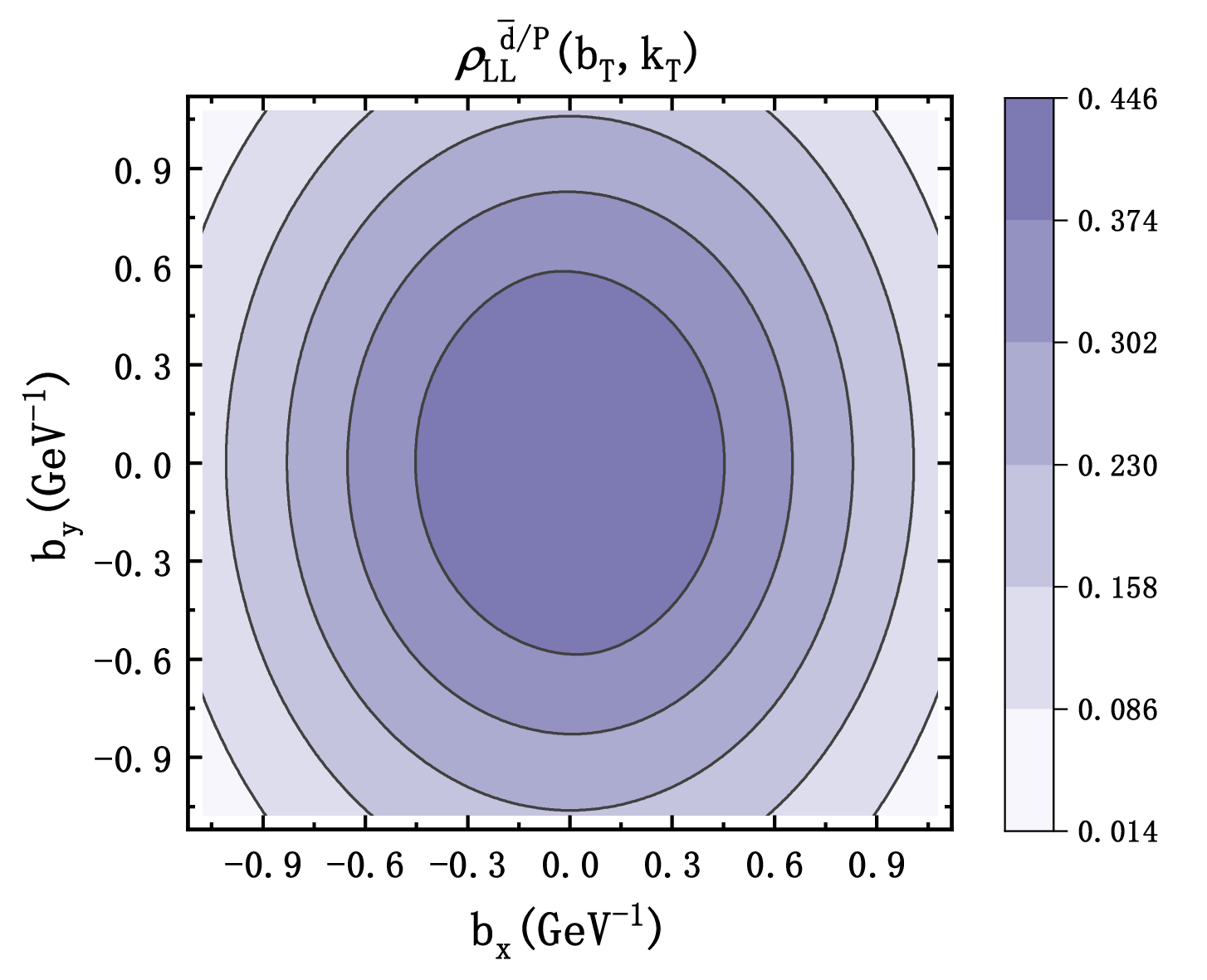}  \end{minipage}}
		\subfigure{\begin{minipage}[b]{0.4\linewidth}
				\centering
				\includegraphics[width=\linewidth]{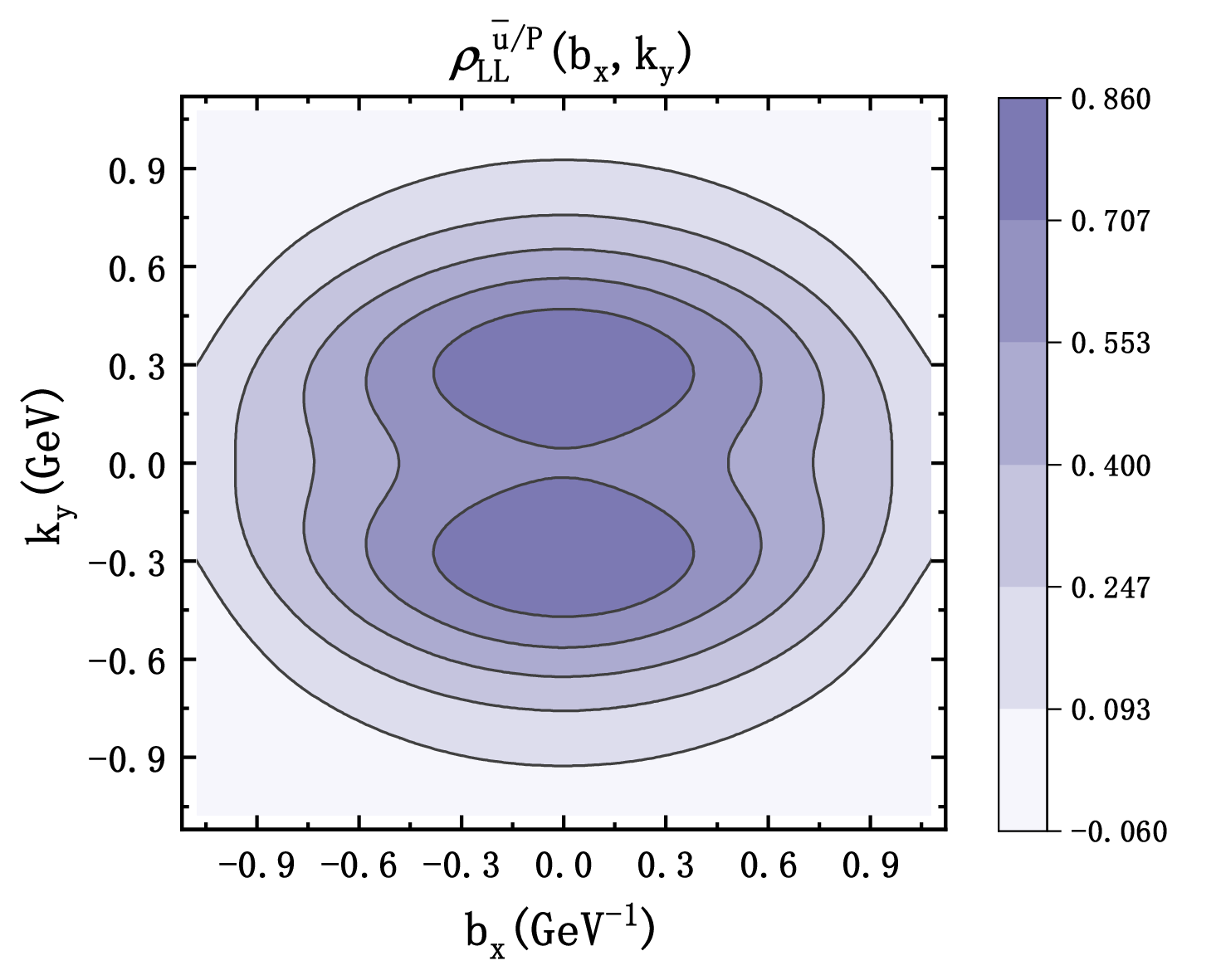}
		\end{minipage}}
		\subfigure{\begin{minipage}[b]{0.4\linewidth}
				\centering
				\includegraphics[width=\linewidth]{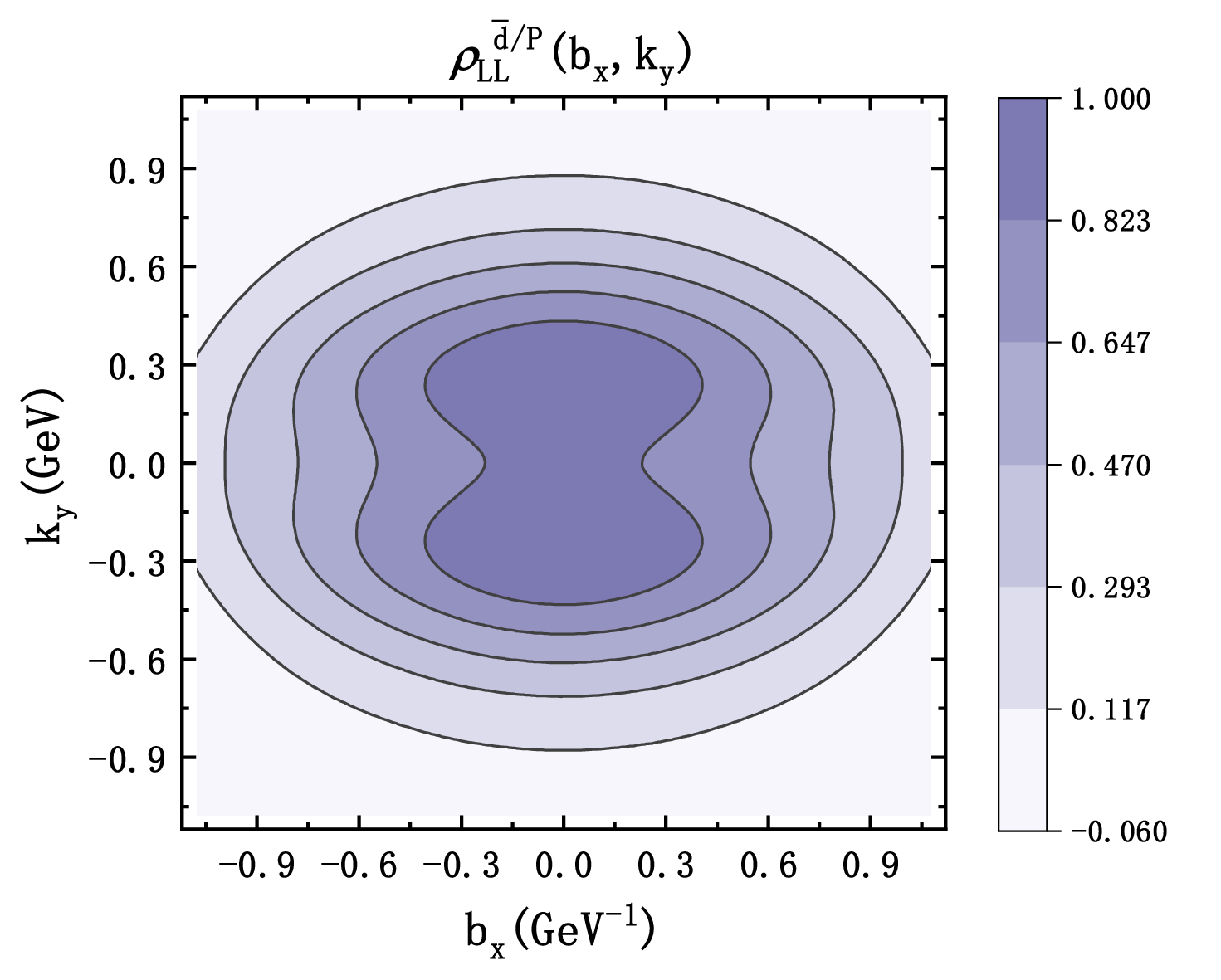}    	\end{minipage}}
		\caption{Similar to Fig.~\ref{uu}, but for the Wigner distribution $\rho_{LL}$ of the $\bar{u}$ (left panel) and $\bar{d}$ (right panel) quarks in the proton.} \label{ll}      
	\end{figure*}

	In this section, we present the numerical results for the Wigner distributions of the sea quarks $\bar{u}$ and $\bar{d}$. 
For the parameters $g_1$, $g_2$, $\Lambda_{\bar{q}}$, $\Lambda_\pi$ in our model, we adopt the values shown in Table.~\ref{tab1} from Ref.~\cite{Luan:2022fjc}, where $g_2$ and $\Lambda_\pi$ are fixed by adopting the GRV leading-order (LO) parametrization~\cite{Gluck:1991ey} to perform the fit for $f_1^{\bar{u}/\pi^-}$ (or $f_1^{\bar{d}/\pi^+}(x)$). The MSTW2008 LO parametrization~\cite{Martin:2009iq} is adopted for $f_1^{\bar{u}/P}$ and $f_1^{\bar{d}/P}$ to obtain the values of the parameters $g_1$ and $\Lambda_{\bar{q}}$.
	\begin{center}\label{tab1}
		\setlength{\tabcolsep}{5mm}
		\renewcommand\arraystretch{1.5}
		\begin{tabular}{ c | c | c }
			\hline
			Parameters & $\bar{u}$ & $\bar{d}$ \\
			\hline
			\hline
			$g_1$ & 9.33 & 5.79 \\
			\hline
			$g_2$ & 4.46 & 4.46 \\
			\hline
			$\Lambda_\pi(GeV)$ &  0.223 & 0.223 \\
			\hline
			$\Lambda_{\bar{q}}(GeV)$ &  0.510 &  0.510 \\
			\hline
		\end{tabular}
		\captionof{table}{Values of the parameters obtained from Ref. \cite{Luan:2022fjc}.} \label{tab1}
	\end{center}
	
The Wigner distribution is a five-dimensional function of $b_x$, $b_y$, $k_x$, $k_y$, $x$. 
Here we only discuss the case of Wigner distributions in the transverse space, namely the transverse-coordinate space (the impact-parameter space ) and the transverse-momentum space.
The transverse Wigner distributions can be obtained by integrating out the longitudinal momentum fraction $x$
\begin{align}
		\rho\left(\boldsymbol{b}_{T}, \boldsymbol{k}_{T}\right) & = \int_{0}^{1} d x \rho\left(x, \boldsymbol{b}_{T}, \boldsymbol{k}_{T}\right),
\end{align}
To extract more information from the Wigner distributions, we also study the mixed transverse Wigner distributions $\rho\left(b_x, k_y\right)$, which is a probability density obtained by integrating over $b_y$ and $k_x$
\begin{align}
		\rho\left(b_{x}, k_{y}\right)=\int d b_{y} \int d k_{x} \rho\left(\boldsymbol{b}_{T}, \boldsymbol{k}_{T}\right).
\end{align}
	
In Fig.~\ref{uu}, We plot contour plots of the $x$-integrated Wigner distribution $\rho_{UU}$ of the $\bar{u}$ (left panel) and $\bar{d}$ (right panel) quarks in the proton. 
The upper panel depicts the distributions in the transverse momentum space with fixed impact parameter $\boldsymbol{b}_{T}=0.3$ GeV$^{-1}\ \hat{\boldsymbol{e}}_{y}$. 
The central panel depicts the distributions in the impact parameter space with fixed transverse momentum $\boldsymbol{k}_{T}=0.3$ GeV $\hat{\boldsymbol{e}}_{y}$.
The lower panel depicts the distributions in the mixed plane.

We can observe a left-right symmetry in Fig.~\ref{uu} which implies that the unpolarized sea quarks in unpolarized proton have no preference for clockwise and anticlockwise motion. 
Comparing the behaviors of the $\bar{u}$ and $\bar{d}$ quarks, we find that  $\rho_{UU}^{\bar{u}/P}$ and $\rho_{UU}^{\bar{d}/P}$ in our model have positive maxima at the center ($k_x=k_y=0$), ($b_x=b_y=0$), ($b_x=k_y=0$) in the transverse momentum plane, transverse coordinate plane and the mixed plane. 
We also note that the spread behaviors of the distributions for $\bar{u}$ and $\bar{d}$ quarks are similar in $\boldsymbol{k}_{T}$ and $\boldsymbol{b}_{T}$ space, that is, the distributions increase faster in $x$-direction than in $y$-direction. 
While we can clearly see that the probability density $\rho_{UU}^{\bar{u}/P}\left(b_x, k_y\right)$ and $\rho_{UU}^{\bar{d}/P}\left(b_x, k_y\right)$ are extended more in $b_x$ than in $k_y$. 

We also calculate the average quadrupole distortions $Q_{b}^{i j}\left(\mathbf{k}_{T}\right)$ and $Q_{k}^{i j}\left(\mathbf{b}_{T}\right)$ defined as follows
	\begin{align}
		\notag 	Q_{b}^{i j}\left(\mathbf{k}_{T}\right)& = \frac{\int    d^{2} \mathbf{b}_{T}\left(2 b_{T}^{i} b_{T}^{j}-\delta^{i j} \mathbf{b}_{T}^{2}\right) \rho_{U U}\left(\mathbf{b}_{T}, \mathbf{k}_{T}\right)}{\int d^{2} \mathbf{b}_{T} \mathbf{b}_{T}^{2} \rho_{U U}\left(\mathbf{b}_{T}, \mathbf{k}_{T}\right)} \\
		Q_{k}^{i j}\left(\mathbf{b}_{T}\right)& = \frac{\int d^{2} \mathbf{k}_{T}\left(2 k_{T}^{i} k_{T}^{j}-\delta^{i j} \mathbf{k}_{T}^{2}\right) \rho_{U U}\left(\mathbf{b}_{T}, \mathbf{k}_{T}\right)}{\int d^{2} \mathbf{k}_{T} \mathbf{k}_{T}^{2} \rho_{U U}\left(\mathbf{b}_{T}, \mathbf{k}_{T}\right)}.
	\end{align}
They can be used to quantitatively estimate the distortion of the unpolarized sea quarks in the nucleon. 
Numerical calculation shows that, at $k_T=0.3 GeV$, $Q_{b}\left(\mathbf{k}_{T}\right)=0.442$ for the $\bar{u}$ quark and 0.307 for the $\bar{d}$ quark, respectively. 
While at $b_T=0.3 GeV^{-1}$, $Q_{k}\left(\mathbf{b}_{T}\right)=-1.535$ for the  $\bar{u}$ quark and $-1.326$ for the $\bar{d}$ quark.  
Therefore, the distributions $\rho_{UU}^{\bar{u}/P}$ and $\rho_{UU}^{\bar{d}/P}$ in transverse momentum plane as well as transverse impact parameter plane have distortions, which indicates the configuration $\boldsymbol{b}_{T} \perp \boldsymbol{k}_{T}$ is favored rather than the configuration $\boldsymbol{b}_{T} \| \boldsymbol{k}_{T}$. 
These distortions are similar to the valence quark results calculated using the light-cone constituent quark model(LCCQM)~\cite{Lorce:2011ni} and chiral quark soliton model ($\chi \mathrm{QSM}$)~\cite{Lorce:2011kd}. 
Our results are different form the results in a light front quark-diquark model where the light-front wave functions are modeled from the soft-wall AdS/QCD prediction.~\cite{Chakrabarti:2017teq}.
		
In Fig.~\ref{lu},  We show the contour plots of the longitudinal-unpolarized Wigner distribution $\rho_{LU}$ for the $\bar{u}$ (left panel) and $\bar{d}$ (right panel) quarks. 
The upper panel shows the plots in the $\bm{k}_T$-space with fixed impact parameter $\boldsymbol{b}_{T}=0.3$ GeV$^{-1} \hat{\boldsymbol{e}}_{y}$, the central panel shows the plots in the $\bm{b}_T$-space with fixed transverse momentum $\boldsymbol{k}_{T}=0.3$ GeV $\hat{\boldsymbol{e}}_{y}$, respectively. 
There are dipole structures of $\bar{u}$ and $\bar{d}$ quarks in the $k_T$-space or in the $b_T$-space.
The signs of the polarities for $\rho_{LU}^{\bar{u}/P}$ and $\rho_{LU}^{\bar{d}/P}$ are the same in the $k_T$-space or in the $b_T$-space. 
Particularly, the distributions are positive in $k_x<0$ region and are negative in $k_x>0$ region, while they are negative in $b_x<0$ region and are positive in $b_x>0$ region. 
The lower panel shows the mixed Wigner distribution $\rho_{LU}^{\bar{q}/P}\left(b_x, k_y\right)$ and depicts a quadrupole structure for both the $\bar{u}$ and $\bar{d}$ quarks. 
These multipole structures are due to the explicit factor $\epsilon^{ij}_Tk_T^i\frac{\partial}{\partial b_T^j}$ in Eq.~(\ref{F14}) which breaks the left-right symmetry and implies that the net OAM of the sea quark is non-zero. 
The average sea quark OAM in a proton polarized in the z-direction can be written as 
	\begin{align}
		\notag \ell_{z}^{\bar{q}/P} & = \int d x d^{2} \boldsymbol{k}_T d^{2} \boldsymbol{b}_T\left(\boldsymbol{b}_T \times \boldsymbol{k}_T\right)_{z} \rho^{\left[\gamma^{+}\right] \bar{q}/P}\left(x, \boldsymbol{b}_T, \boldsymbol{k}_T\right) \\
		&\notag = \int d x d^{2} \boldsymbol{k}_T d^{2} \boldsymbol{b}_T\left(\boldsymbol{b}_T \times \boldsymbol{k}_T\right)_{z}
		\\&\times\left[\rho_{U U}^{\bar{q}/P}\left(x,\boldsymbol{b}_T, \boldsymbol{k}_T\right)+\rho_{L U}^{\bar{q}/P}\left(x,\boldsymbol{b}_T, \boldsymbol{k}_T\right)\right].
	\end{align}
If $\boldsymbol{k}_T$ and $\boldsymbol{b}_T$ are integrated out for $\rho_{U U}^{\bar{q}/P}$, the result is zero, which means that the total angular momentum of constituents sum up to zero in  an unpolarized nucleon. 
Then the sea quark OAM can be written in terms of GTMDs as
	\begin{align}
		\notag \ell_{z}^{\bar{q}/P}& = \int d x d^{2} \boldsymbol{k}_T d^{2} \boldsymbol{b}_T\left(\boldsymbol{b}_T \times \boldsymbol{k}_T\right)_{z}\rho_{L U}^{\bar{q}/P}\left(x,\boldsymbol{b}_T, \boldsymbol{k}_T\right)\\
		&= -\int d x d^{2} \boldsymbol{k}_T \frac{\boldsymbol{k}_T^{2}}{M^{2}} F_{14}^{\bar{q}/P}\left(x, 0, \boldsymbol{k}_T^{2}, 0,0\right).
	\end{align}
Numerical calculation yields $\ell_{z}^{\bar{u}/P}=0.027$ and $\ell_{z}^{\bar{d}/P}=0.051$,
which are positive for both the $\bar{u}$ and $\bar{d}$ quarks. 
This indicates the sea quark OAM is parallel to the proton spin for both the $\bar{u}$ and $\bar{d}$ quarks in our model.
	
In Fig.~\ref{ul},  We show the contour plots of the unpolarized-longitudinal Wigner distribution $\rho_{UL}\left(\boldsymbol{b}_{T}, \boldsymbol{k}_{T}\right)$ and the mixed distributions $\rho_{UL}\left(b_x, k_y\right)$ for the $\bar{u}$ (left panel) and $\bar{d}$ (right panel) quarks.  
Similar to the longitudinal-unpolarized Wigner distribution, $\rho_{UL}\left(\boldsymbol{b}_{T}, \boldsymbol{k}_{T}\right)$ also has dipolar structures in both the $k_T$-space and in the $b_T$-space.
However, the sign of $\rho_{UL}$ is opposite to that of $\rho_{LU}$.  
In the lower panel, we can observe that the mixed Wigner distribution $\rho_{UL}^{\bar{q}/P}\left(b_x, k_y\right)$ has a quadrupole structure.
Again, the multipole structure is due to the factor $\epsilon^{ij}_T 
\ k_T^i\frac{\partial}{\partial b_T^j}$ in Eq.~(\ref{G14}) which essentially reflects quark spin-orbit correlations. 
The correlation between the longitudinal spin and the OAM of the sea quarks can be defined as 
\begin{align}
		\notag	C_{z}^{\bar{q}/P} & = \int d x d^{2} \boldsymbol{k}_{T} d^{2} \boldsymbol{b}_{T}\left(\mathbf{b}_{\perp} \times \boldsymbol{k}_{T}\right)_{z} \rho_{U L}^{\bar{q}/P}\left(x,\boldsymbol{k}_{T}, \boldsymbol{b}_{T}\right) \\
		& = \int d x d^{2} \boldsymbol{k}_{T} \frac{\boldsymbol{k}_{T}^{2}}{M^{2}} G_{11}^{\bar{q}/P}\left(x, 0, \boldsymbol{k}_{T}^{2}, 0,0\right).
\end{align}
Numerical calculation shows that both the sea quarks have negative spin-orbit correlation $C_{z}^{\bar{u}/P}=-0.245$ ang $C_{z}^{\bar{d}/P}=-0.394$, which implies the sea quark spin and OAM tend to be anti-aligned.
	
In Fig.~\ref{ll},  We plot the longitudinal-longitudinal Wigner distributions $\rho_{LL}$ for the $\bar{u}$ (left panel) and $\bar{d}$ (right panel) quarks in a way similar to Fig.~\ref{uu}.
These distributions describe the phase-space distributions of longitudinal polarized quark in a longitudinal polarized proton, and correspond to the axial charge $(\Delta q)$ of the nucleon after integrating over transverse variables. 
We find the distributions of both $\bar{u}$ and $\bar{d}$ quarks are positive, which implies the signs of $\Delta\bar{u}$ and $\Delta\bar{d}$ are positive. 
For the mixed Wigner distribution, a sign change is observed in large $b_x$ or $k_y$. 
This kind of sign change is also found in the longitudinal-longitudinal Wigner distributions of quarks~\cite{Lorce:2011kd,Liu:2015eqa}.

\section{CONCLUSION}\label{Sec4} 
	
In this work, we studied the five-dimensional Wigner distributions of sea quarks in the proton using a light-cone model, in which the Wigner distributions and the GTMDs can be expressed as the overlap of LCWFs. 
To generate the sea quark degree of freedom, we treated the Fock state of proton as a composite system formed by a pion meson and a baryon, where the pion meson is composed in terms of $q\bar{q}$. 
We numerically calculated the four Wigner distributions $\rho_{UU}$, $\rho_{LU}$, $\rho_{UL}$, and  $\rho_{LL}$ of the $\bar{u}$ and $\bar{d}$ quarks in the transverse momentum space and the transverse position space within the overlap representation.
Distortions can be found in the distributions $\rho_{LU}$ and $\rho_{UL}$ of the $\bar{u}$ and $\bar{d}$ quarks. 
Particularly, the dipole structures have been observed in $\rho_{LU}\left(\boldsymbol{b}_{T}, \boldsymbol{k}_{T}\right)$ and $\rho_{UL}\left(\boldsymbol{b}_{T}, \boldsymbol{k}_{T}\right)$; and the quadrupole structures have been observed in the mixing distributions $\rho_{LU}\left(\boldsymbol{b}_{x}, \boldsymbol{k}_{y}\right)$ and $\rho_{UL}\left(\boldsymbol{b}_{x}, \boldsymbol{k}_{y}\right)$.
The polarities of the multi-pole structures in $\rho_{LU}$ are opposite to those in the $\rho_{UL}$. 
The result of the averaged quadrupole distortions $Q_{b}^{i j}\left(\mathbf{k}_{T}\right)$ and $Q_{k}^{i j}\left(\mathbf{b}_{T}\right)$ 
indicates that the configuration $\boldsymbol{b}_{T} \perp \boldsymbol{k}_{T}$ is favored rather than $\boldsymbol{b}_{T} \| \boldsymbol{k}_{T}$, which is similar to the results for the valence quarks calculated from the light-cone constituent quark model and chiral quark soliton model.  
We also evaluated the spin-orbit correlation $C_{z}$ and the OAM of the $\bar{u}$ and $\bar{d}$ quarks using the relation between Wigner distributions and GTMDs. 
The study on sea quark Wigner distributions may provide useful information about the sea quarks in proton as well as improve our understanding on the multidimensional image of the proton in the quantum phase space.

\section*{Acknowledgements}
This work is partially supported by the National Natural Science Foundation of China under grant number 12150013.

\end{document}